\documentclass[a4paper,fleqn]{cas-dc}

\usepackage[numbers]{natbib}
\usepackage{url}
\usepackage[utf8]{inputenc}
\usepackage{caption}
\usepackage{times} 
\usepackage{amssymb}
\usepackage{graphicx}
\usepackage{amsmath}
\usepackage{amsthm}
\usepackage{booktabs}
\usepackage{algorithm}
\usepackage{algorithmic}
\usepackage{xcolor}  
\usepackage{multirow}
\usepackage{makecell}
\usepackage{array}
\usepackage{tikz}
\usetikzlibrary{shapes.geometric, arrows.meta, positioning}
\usepackage{makecell}
\usepackage[switch]{lineno}

\newcommand\mingwang[1]{\textit{\textcolor{cyan}{[mingwang: #1]}}}
\newcommand\wanglj[1]{\textit{\textcolor{blue}{[LJ: #1]}}}
\newcommand\todo[1]{\textit{\textcolor{red}{[todo: #1]}}}

\definecolor{rebbtal}{RGB}{46,117,183}  

\usepackage{enumitem}
\setlist[itemize]{align=parleft,left=0pt..1em}

\def\tsc#1{\csdef{#1}{\textsc{\lowercase{#1}}\xspace}}
\tsc{WGM}
\tsc{QE}

\UseRawInputEncoding
\begin{document}
\let\WriteBookmarks\relax
\def\floatpagepagefraction{1}
\def\textpagefraction{.001}

\shorttitle{}    

\shortauthors{}  

\title [mode = title]{Adversarial Attacks and Defenses on Text-to-Image Diffusion Models: A Survey}  



%

\author[1]{Chenyu Zhang}




\author[1]{Mingwang Hu}
\author[1]{Wenhui Li}
\author[1]{Lanjun Wang}
\cormark[1]
\ead{wanglanjun@tju.edu.cn}


\affiliation[1]{organization={Tianjin University},
            city={Tianjin},
            country={China}}






\cortext[1]{Corresponding author}



\begin{abstract}
Recently, the text-to-image diffusion model has gained considerable attention from the community due to its exceptional image generation capability. 
A representative model, Stable Diffusion, amassed more than 10 million users within just two months of its release.
This surge in popularity has facilitated studies on the robustness and safety of the model, leading to the proposal of various adversarial attack methods. 
Simultaneously, there has been a marked increase in research focused on defense methods to improve the robustness and safety of these models.
In this survey, we provide a comprehensive review of the literature on adversarial attacks and defenses targeting text-to-image diffusion models.
We begin with an overview of text-to-image diffusion models, followed by an introduction to a taxonomy of adversarial attacks and an in-depth review of existing attack methods.
We then present a detailed analysis of current defense methods that improve model robustness and safety.
Finally, we discuss ongoing challenges and explore promising future research directions.
For a complete list of the adversarial attack and defense methods covered in this survey, please refer to our curated repository at \url{https://github.com/datar001/Awesome-AD-on-T2IDM}.
\textcolor{red}{\textbf{Warning:} This paper includes model-generated content that may contain offensive or distressing material.}
\end{abstract}





\maketitle

\section{Introduction}
\label{sec:introduction}
The text-to-image model aims to generate a range of images based on user-provided prompt. Recent advances in deep learning have led to the proposal of numerous text-to-image models~\cite{DBLP:conf/icml/RameshPGGVRCS21, yu2022scaling, DBLP:journals/corr/abs-2105-13290, ding2022cogview2, DBLP:journals/corr/abs-2112-10741, DBLP:journals/corr/abs-2111-12417, DBLP:conf/nips/SahariaCSLWDGLA22, BetkerImprovingIG}, significantly enhancing the quality of generated images. 
As a prominent branch of text-to-image models, the diffusion model has emerged as a research hotspot owing to its superior image generation capabilities~\cite{DBLP:conf/cvpr/RombachBLEO22, Midjourney, ho2020denoising, saharia2022photorealistic, ruiz2023dreambooth, brooks2023instructpix2pix, peebles2023scalable, nichol2021glide, yi2024diff, huang2024stfdiff}.
The representative diffusion model, Stable Diffusion~\cite{DBLP:conf/cvpr/RombachBLEO22} and Midjourney~\cite{Midjourney}, boast user bases that exceed 10 million~\cite{stablediffusionstatistics} and 14.5 million~\cite{midjourystatistics}, indicating that text-to-image generation has become an integral part of daily life.

Despite their advances, the text-to-image diffusion model exhibits vulnerabilities in both \textbf{robustness} and \textbf{safety}. 
Robustness ensures that the model can generate images with consistent semantics in response to various input prompts in real-world scenarios.
Safety prevents misuse of the model in creating malicious images, such as sexual, violent, and politically sensitive images, etc.

For the robustness of the model, existing attack methods uncover two types of vulnerability: 1) The model often generates inaccurate images against the prompt with multiple objects and attributes. 
Hila et al.~\cite{chefer2023attend} 
highlight that Stable Diffusion faces challenges in generating multiple objects from a single prompt and accurately associating properties with the correct objects. Similarly, Du et al.~\cite{du2023stable} note that similarity in the appearance and generation speed of different objects affect the quality of the generated images. 
2) The model lacks robustness to the grammatically incorrect prompt with subtle noise. 
Zhuang et al.~\cite{10208563} demonstrate that inserting just five random characters into a prompt can significantly alter the content of the output images. 
Moreover, some typos, glyph alterations, or phonetic changes in the prompt can also destroy the semantics of the image~\cite{gao2023evaluating}.

For the safety issues of the model, 
UnsafeDiffusion~\cite{qu2023unsafe} assesses the safety of both open-source and commercial text-to-image diffusion models, uncovering their potential to generate images that are sexually explicit, violent, disturbing, hateful, or politically sensitive. 
Although various safeguards~\cite{midjoury-safety, DALL-E_3_System_Card, nsfw_list, image_filter_1, image_filter_2} are implemented within the model to mitigate the output of malicious content, several adversarial attack methods~\cite{yang2023sneakyprompt, ba2023surrogateprompt, zhang2023generate, Yang2023MMADiffusionMA, deng2023divideandconquer} have been proposed to craft the adversarial prompt, which circumvents these safeguards and effectively generates malicious images, posing significant safety threats.

Exposure to vulnerabilities in the robustness and safety of text-to-image models underscores the urgent need for effective defense mechanisms. 
For the vulnerability in the robustness of the model, many works~\cite{lian2023llm, bar2023multidiffusion, podell2023sdxl, avrahami2023spatext} are proposed to improve the generation capability against the prompt with multiple objects and attributes. 
However, the robustness improvement of the model against the grammatically incorrect prompt with subtle noise remains underexplored.
For the vulnerability in the safety of the model, existing defense methods can be divided into two types: external safeguards and internal safeguards.
External safeguards~\cite{liu2024latent, wu2024universal, yang2024guardt2i} focus on detecting or correcting the malicious prompt before feeding the prompt into the text-to-image model.
In contrast, internal safeguards~\cite{gandikota2023erasing, kumari2023ablating, gandikota2023unified, orgad2023editing, Schramowski2022SafeLD, li2023self} aim to ensure that the semantics of output images deviate from those of malicious images by modifying internal parameters and features within the model.

Despite these efforts, significant limitations and challenges persist in the field of adversarial attacks and defenses. 
A primary issue with existing attack methods is their lack of imperceptibility. Specifically, most attack methods~\cite{10208563, yang2023sneakyprompt, zhang2023generate, Yang2023MMADiffusionMA} 
focus on adding a noise word or rewriting the entire prompt to craft the adversarial prompt. This significant added noise can be easily detected, thus reducing the attack imperceptibility.
Another challenge is that current defense methods are not adequately equipped to deal with the threat posed by adversarial attacks. 
Specifically, existing methods for improving robustness show a marked deficiency in countering grammatically incorrect prompts with subtle noise~\cite{10208563, gao2023evaluating}. 
Moreover, while existing defense methods~\cite{wu2024universal, gandikota2023erasing, kumari2023ablating, orgad2023editing, Schramowski2022SafeLD, li2023self} to improve safety are somewhat effective against malicious prompts that explicitly contain malicious words~(such as nudity), they are less successful against adversarial prompts that cleverly omit malicious words.



\textbf{Related Survey.} Although there have been innovative contributions in this field, a comprehensive review work that systematically collates and synthesizes these efforts is currently lacking.
Existing surveys on adversarial attacks and defenses mainly focus on natural language processing~\cite{10.1145/3374217, 9587104, 10.1145/3593042, Shayegani2023SurveyOV}, computer vision~\cite{8294186, 9614158}, explainable artificial intelligence~\cite{baniecki2024adversarial}, federated learning~\cite{rodriguez2023survey}, etc. Moreover, existing surveys on the text-to-image diffusion model focus on the technology development~\cite{zhang2023texttoimage, yang2023diffusion, zelaszczyk2024text, croitoru2023diffusion}, image quality evaluation~\cite{hartwig2024evaluating}, controllable generation~\cite{cao2024controllable}, etc.
In contrast, our work is the first survey dedicated to providing an in-depth review of adversarial attacks and defenses on the text-to-image diffusion model~(AD-on-T2IDM). 

\textbf{Paper Selection.} Given that AD-on-T2IDM is an emerging field, we review high-quality papers from leading AI conferences such as CVPR\footnote{CVPR: IEEE Conference on Computer Vision and Pattern Recognition}, ICCV\footnote{ICCV: IEEE International Conference on Computer Vision}, NeurIPS\footnote{NeurIPS: Conference on Neural Information Processing Systems}, AAAI\footnote{AAAI: Association for the Advancement of Artificial Intelligence Conference}, ICLR\footnote{ICLR: International Conference on Learning Representations}, ACM MM\footnote{ACM MM: ACM International Conference on Multimedia}, WACV\footnote{WACV: Winter Conference on Applications of Computer Vision}, and prominent information security conferences like SP\footnote{SP: IEEE Symposium on Security and Privacy} and CCS\footnote{CCS: ACM Conference on Computer and Communications Security}. In addition to accepted papers at these conferences, we also consider notable submissions from the e-Print archive\footnote{\url{https://arxiv.org/}}, which represent the cutting edge of current research. We manually selected papers submitted to the archive before June 1, 2024, using two criteria: paper quality and methodology novelty.

\textbf{Paper Structure.} In this work, we begin with an overview of text-to-image diffusion models in Sec.~\ref{sec-text-to-image-model}. Subsequently, we review adversarial attacks and defenses on the text-to-image model in Sec.~\ref{sec:attack} and Sec.~\ref{sec: defense}, respectively. Next, we analyze the limitations of existing attack and defense methods and discuss the potential solution in Sec.~\ref{sec: challenge}.
Finally, we conclude this survey in Sec.~\ref{sec:conclusion}.

\section{Text-to-Image Diffusion Model}
\label{sec-text-to-image-model}
Following the previous survey study~\cite{zhang2023texttoimage}, existing diffusion models can be roughly divided into two categories: diffusion model in pixel space and in latent space. Representative models in pixel space include GLIDE~\cite{nichol2021glide}, Imagen~\cite{DBLP:conf/nips/SahariaCSLWDGLA22}, while those in latent space include Stable Diffusion~\cite{DBLP:conf/cvpr/RombachBLEO22} and DALL$\cdot$E~2~\cite{dalle2}. In particular, the open-source Stable Diffusion has been extensively deployed across various platforms, such as ClipDrop~\cite{clipdrop}, Discord~\cite{discord}, and DreamStudio~\cite{sdxl}, and has generated over 12 billion images, which represents 80\% of AI images on the Internet~\cite{stablediffusionstatistics}. Consequently, most adversarial attacks~\cite{du2023stable, 10208563, yang2023sneakyprompt, Yang2023MMADiffusionMA, maus2023black, shahgir2023asymmetric, zhang2024revealing, 10205174} and defenses~\cite{gandikota2023erasing, kumari2023ablating, gandikota2023unified, orgad2023editing, kim2023safe, arad2023refact, poppi2024safe} are based on Stable Diffusion. In the following sections, we introduce the framework of Stable Diffusion. For clarity, we define the notations used in this work in Table~\ref{tab:symbol}.

Stable Diffusion is composed of three primary modules: the image autoencoder, the conditioning network, and the denoising network. Firstly, the image autoencoder is pre-rained on a diverse image dataset to achieve the transformation between the pixel space and the latent space. Specifically, the image autoencoder contains an encoder and a decoder, where the encoder aims to transform an input image $y$ into a latent representation $z = \mathcal{E}(y)$ and the decoder conversely reconstructs the input image from the latent representation, i.e., $\mathcal{D}(z) = \hat{y} \approx y$.
For the conditioning network, Stable Diffusion utilizes a pre-trained text encoder, CLIP~\cite{clip}, to encode the input prompt $x$ into the prompt feature $c = E_{txt}(x)$, which serves as a text condition to guide the image generation process.
The denoising network is a UNet-based diffusion model $\epsilon_{\theta}$ to craft the latent representation guided by a text condition. 
In the following section, we introduce the training and inference stages of Stable Diffusion, respectively.

\textit{Training Stage}.
Following a pioneering work, DDPM \cite{ho2020denoising}, the training of Stable Diffusion can be seen as a Markov process in the latent space that iteratively adds Gaussian noise to the latent representation of the image during a forward process and then restores the latent representation through a reverse denoising process.

The forward process aims to iteratively add Gaussian noise to the latent representation of the image until it reaches a state of random noise. Give an image $y$, Stable Diffusion first employs the pre-trained encoder $\mathcal{E}$ to obtain the initial latent representation $z_0 = \mathcal{E}(y)$. Then, a diffusion process is conducted by adding Gaussian noise to $z_0$ as follows:
\begin{equation}
    q(z_t|z_{t-1}) = \mathcal{N}(z_t;\sqrt{1-\beta_t}z_{t-1}, \beta_t\mathcal{I}), t \in (0, T)
\end{equation}
where $\beta_t$ is the pre-defined hyper-parameter, $t$ and $T$ is the current and total diffusion step. 
With $\alpha_t=1-\beta_t$ and $\bar{\alpha}_t=\prod_{i=1}^t \alpha_i$, the noisy latent representation $z_t$ at timestep $t$ can be obtained as:
\begin{equation}
    q(z_t|z_0) = \mathcal{N}(z_t; \sqrt{\bar{\alpha}_t}z_0, (1-\sqrt{\bar{\alpha}_t})\mathcal{I}).
\end{equation}

The reverse process aims to gradually remove the noise in the noisy latent representation until it reaches the initial latent representation $z_0$. 
Therefore, the model predicts the latent noise~(Gaussian noise) $\epsilon_t$ added to the latent representation conditioned on the time step $t$ and the text condition $c$.
The training objective is encapsulated by the loss function:
\begin{equation}
    \mathcal{L} = \mathbb{E}_{\epsilon \in \mathcal{N}(0,1), z, c, t}[|| \epsilon_t - \epsilon_{\theta}(z_t, c, t) ||_2^{2}].
\end{equation}

\textit{Inference Stage}.
Classifier-free guidance~\cite{ho2022classifier} is used to regulate image generation in the inference stage. Specifically, the text-to-image diffusion model first predicts the latent noise of conditional and unconditional terms, respectively. Then, the final latent noise is directed towards the latent noise of the conditional term and away from that of the unconditional term by utilizing a guidance scale $\alpha > 1$:
\begin{equation}
    \widetilde{\epsilon}_{\theta}(z_t, c, t) = \epsilon_{\theta}(z_t, \phi, t) + \alpha(\epsilon_{\theta}(z_t, c, t) - \epsilon_{\theta}(z_t, \phi, t))
    \label{eq-sd}
\end{equation}
where $\epsilon_{\theta}(z_t, c, t)$ and $\epsilon_{\theta}(z_t, \phi, t)$ are the latent noise of conditional and unconditional terms. The inference process starts from a Gaussian noise $z_T$ and is denoised with $\widetilde{\epsilon}_{\theta}(z_T, c, T)$ to get $z_{T-1}$. This denoising process is done sequentially till $z_0$, which is then transformed to the output image using a decoder $y = D(z_0)$.

\section{Attacks}
\label{sec:attack} 

In this section, we begin by presenting a general framework of adversarial attacks in the text-to-image task.
Then, we propose a classification of existing adversarial attack techniques that are specifically tailored to the given task. 
Finally, expanding on this categorization, we provide an in-depth analysis of the different approaches utilized.


\begin{table}[]
    \centering
    \begin{tabular}{l|l}
        \toprule
        Notation & Definition \\
        \midrule
        $x_{clean}$ & Clean prompt \\
        $y_{clean}$ & Clean image generated by $x_{clean}$ \\
        $x_{mal}^c$ & Malicious concept, such as `nudity' \\
        $x_{mal}$  & Malicious prompt containing $x_{mal}^c$ \\
        $y_{mal}$ & Malicious image generated by $x_{mal}$ \\
        $x_{adv}$ & Adversarial prompt \\
        $y_{adv}$ & Adversarial image generated by $x_{adv}$ \\
        $E_{txt}$ & CLIP text encoder \\
        $E_{img}$ & CLIP image encoder \\
        $\epsilon_{\theta}$ & UNet-based denoising network \\
        \bottomrule
    \end{tabular}
    \caption{Nomenclature}
    \label{tab:symbol}
\end{table}

\noindent \textbf{General Framework.}
The attack framework generally consists of a \textbf{victim model} and a prompt input by the user. Initially, the adversary formulates an \textbf{attack objective} based on the intent. Following this, a \textbf{perturbation strategy} is developed to add noise into the prompt. Finally, based on the attack objective and knowledge of the victim model, the adversary employs an \textbf{optimization strategy} to optimize the noise to craft the final adversarial prompt. 

\begin{figure}
  \centering
   \includegraphics[width=0.9\linewidth]{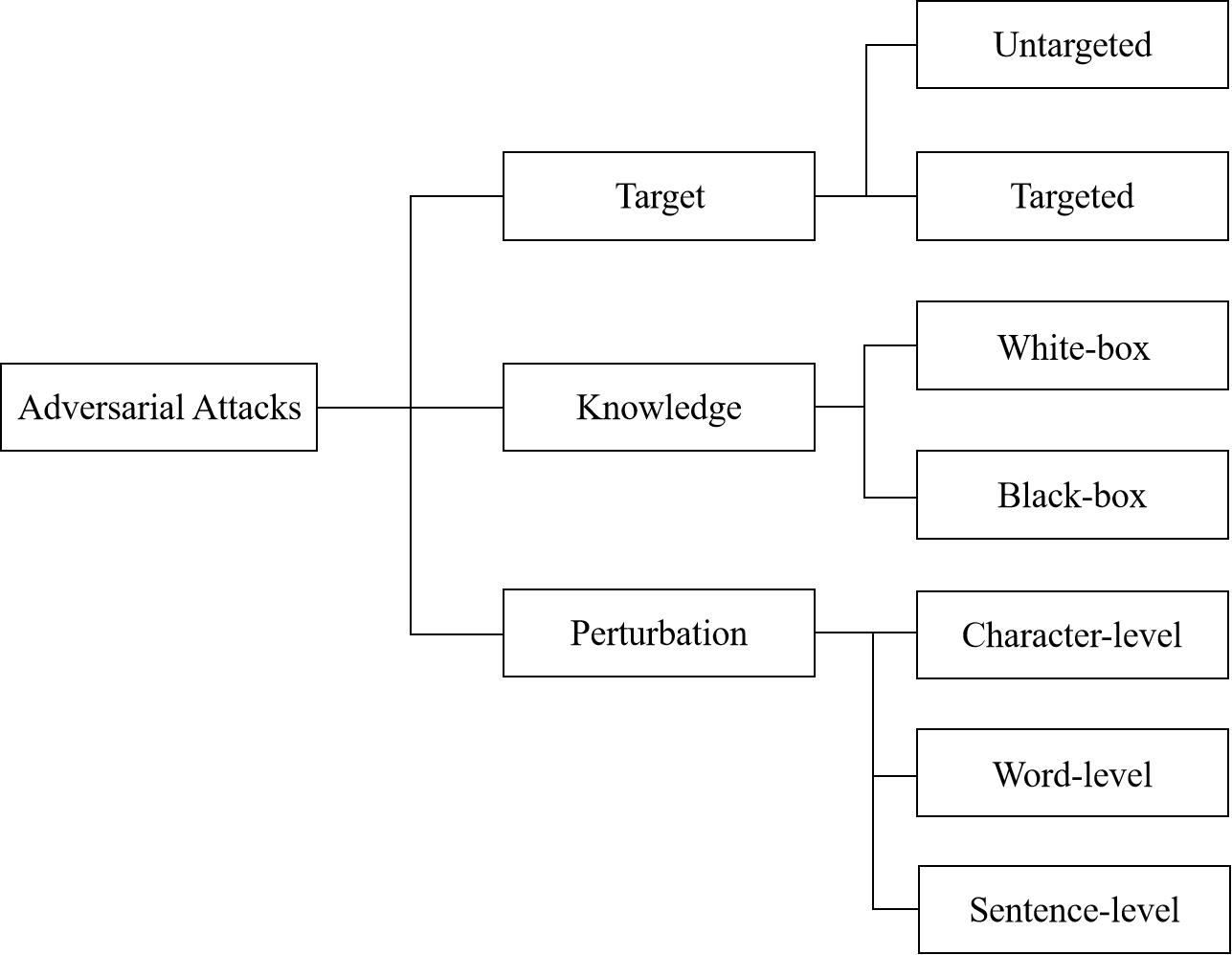}
   \caption{Taxonomy of adversarial attacks on text-to-image models. 
   }
   \label{fig:taxonomy_attack}
\end{figure}

\subsection{Taxonomy of Adversarial Attacks}
\label{subsec:taxonomy_attack}
As shown in Fig.~\ref{fig:taxonomy_attack}, we categorize attack methods based on three aspects: 1) Target or not, 2) how much Knowledge the adversary has, and 3) the type of Perturbation. 

\subsubsection{\textbf{Target}} Based on the intent of the adversary~\cite{10.1145/3374217}, existing attack methods can be divided into two primary categories: untargeted and targeted attacks. 
\begin{itemize}
    \item For untargeted attacks, consider a scenario with a prompt input by the user~(\textbf{clean prompt}) and its corresponding output image~(\textbf{clean image}). The objective of untargeted attacks is to subtly perturb the clean prompt to craft an \textbf{adversarial prompt}, further misleading the victim model to generate an \textbf{adversarial image} with semantics different from the clean image. This type of attack is commonly used to uncover the vulnerability in the robustness of the text-to-image diffusion model.
    \item For targeted attacks, assumes that the victim model has built-in safeguards to filter \textbf{malicious prompts} and resultant \textbf{malicious images}. These prompts and images often explicitly contain \textbf{malicious concepts}, such as `nudity', `violence', and other predefined concepts. The objective of targeted attacks is to obtain an \textbf{adversarial prompt}, which can bypass these safeguards while inducing the victim model to generate \textbf{adversarial images} containing malicious concepts. This type of attack is typically designed to reveal the vulnerability in the safety of the text-to-image diffusion model.
\end{itemize}

\subsubsection{\textbf{Perturbation}} As the adversary is to perturb the prompt in text format, existing attack methods can be divided into three categories according to the perturbation granularity~\cite{10.1145/3593042}: character-level, word-level, and sentence-level perturbations. 
\begin{itemize}
    \item The character-level perturbation involves altering, adding, and removing characters within the word.
    \item The word-level perturbation aims to perturb the word within the prompt, including replacing, inserting, and deleting words within the prompt. 
    \item The sentence-level perturbation alters the sentence within the prompt, which typically involves rewriting the prompt.
\end{itemize}


\subsubsection{\textbf{Knowledge}} Based on the adversary's knowledge of the victim model \cite{10.1145/3374217}, existing methods can be classified as white-box attacks and black-box attacks. 
\begin{itemize}
    \item White-box attacks involve an adversary having full knowledge of the victim model, including its architecture and parameters. Furthermore, based on the model knowledge, the adversary can craft a gradient-based optimization method to learn the adversarial prompt.
    \item Black-box attacks occur when the adversary has no knowledge of the internal working mechanism of the victim model and relies solely on external output to deduce information and create their attacks. For example, Ba~et~al.~\cite{ba2023surrogateprompt} and Deng~et~al.~\cite{deng2023divideandconquer} utilize API access to query the text-to-image diffusion model.
\end{itemize}

\begin{figure*}
    \centering
    \includegraphics[width=1.\linewidth]{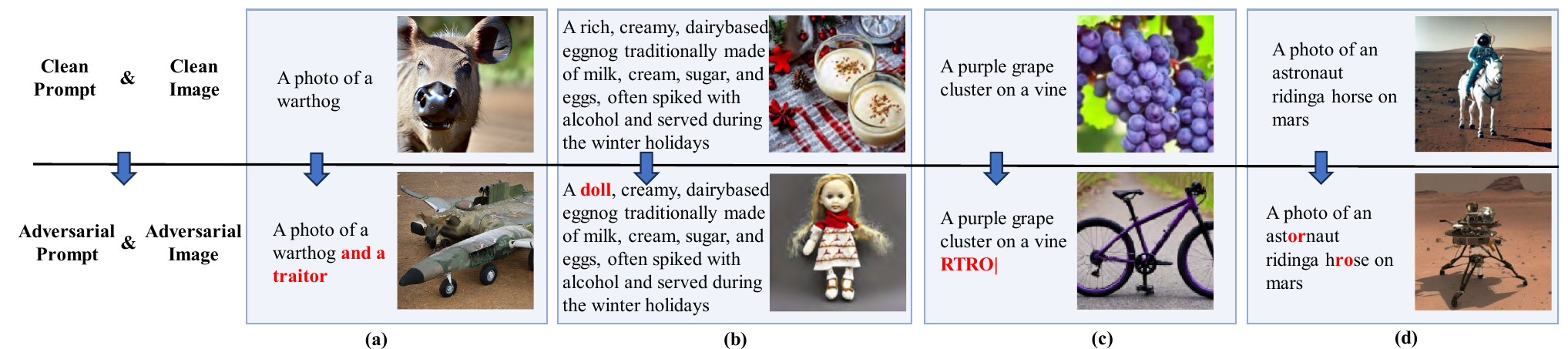}
    \caption{The perturbation strategies of untargeted attack methods. The red words in the adversarial prompt are the noise added to the clean prompt. 
    \textbf{(a)} and \textbf{(b)}, ATM~\cite{du2023stable}, the word-level perturbation by suffix addition and word substitution strategies. \textbf{(c)}, Zhuang~et~al.~\cite{10208563}, the word-level perturbation by appending an noise word with five characters. \textbf{(d)}, Gao~et~al.~\cite{gao2023evaluating}, the character-level perturbation using the typo.
    In these untargeted attacks, the adversarial prompt in \textbf{(a)} is grammatically correct, whereas the prompts in \textbf{(b)}, \textbf{(c)}, and \textbf{(d)} are grammatically incorrect.
    }
    \label{perturbation untargeted attacks}
\end{figure*}

In the following sections, we introduce existing attack methods from two main perspectives: untargeted attacks and targeted attacks.

\subsection{Untargeted Attack}
\label{subsec:untargeted}

This section will analyze three untargeted attack methods to reveal the vulnerability in the robustness of the model, where two of them~\cite{du2023stable, 10208563} are white-box attacks, and the rest one~\cite{gao2023evaluating} focuses on the black-box attack. 

\subsubsection{\textbf{White-box Attacks}}
Early works often explore the robustness of text-to-image diffusion models by manually designed prompts. Typically, Structure Diffusion~\cite{feng2022training} observes that some attributes in the prompt are not correctly aligned with the image content. Additionally, Attend-and-Excite~\cite{chefer2023attend} highlights that Stable Diffusion struggles to generate multiple objects from a prompt
and fails to accurately bind properties to the corresponding objects.
However, manually designing adversarial prompts is time-consuming. To achieve the systematical analysis, two white-box untargeted attack studies~(ATM~\cite{du2023stable} and Zhuang~et~al.~\cite{10208563}) are proposed.  The major similarities of these two studies include: a) both set the popular open-source text-to-image model, Stable Diffusion as the victim model; b) both use Projected Gradient Descent (PGD)~\cite{madry2019deep}, a classical optimization method in the field of adversarial attack, to achieve the attack objective. Despite these similar aspects, these two studies have different attack objectives and perturbation strategies:

\begin{itemize}
    \item \textit{Attack Objective}. 
    ATM~\cite{du2023stable} uses an image classifier $f$ to identify the semantic label of the output image, 
    and maximizes classification loss to deviate the semantics of adversarial images from those of clean images as follows: 
    \begin{equation}
        \max L_{cls}(f(y_{adv}), \arg\max(f(y_{clean}))).
        \label{objective untargeted classification}
    \end{equation}
    where $\arg\max(f(y_{clean}))$ is the label of the clean image~$y_{clean}$ and $L_{cls}$ is the cross-entropy loss.
    However, optimizing Eq.~\ref{objective untargeted classification} is challenging due to the costly computation complexity involved in the model that maps the prompt to the probability distribution of the generated image. To address this problem,
    Zhuang~et~al.~\cite{10208563} transform the semantic deviation of adversarial images into that of adversarial prompts, and minimize the cosine similarity between adversarial and clean prompts in the CLIP embedding space:
    \begin{equation}
        \min cos(E_{txt}(x_{adv}), E_{txt}(x_{clean}))
        \label{objective untargeted cosine}
    \end{equation}
    where $cos(\cdot, \cdot)$ represents the cosine similarity.
    \item \textit{Perturbation Strategy}.
    Although both methods provide word-level perturbation, the concrete perturbation methods are different. As shown in Fig.~\ref{perturbation untargeted attacks}(a) and Fig.~\ref{perturbation untargeted attacks}(b), ATM~\cite{du2023stable} explores two perturbation strategies: word substitution and suffix addition. In contrast, Zhuang~et~al.~\cite{10208563} perturb the clean prompt by appending a noise word with five random characters, as shown in Fig.~\ref{perturbation untargeted attacks}(c).
\end{itemize}
In addition to the above comparison, ATM~\cite{du2023stable} additionally identifies four patterns of adversarial prompts related to vulnerabilities of Stable Diffusion, which further enhance the understanding of researchers about the generation deficit of Stable Diffusion. 

\subsubsection{\textbf{Black-box Attacks}}
\label{subsubsec:black_untargeted}
There is only one black-box untargeted attack~\cite{gao2023evaluating}, which achieves the semantic deviation by maximizing the distribution discrepancy between the adversarial and clean images. Specifically, the discrepancy can be measured by the Maximum Mean Discrepancy (MMD)~\cite{Tolstikhin2016MinimaxEO}: 

{\small
\begin{equation}
\label{objective untargeted MMD}
\begin{aligned}
    \max & \big(\frac{1}{N^2} \sum_{i=1}^N \sum_{j=1}^N k(y_{clean}^i, y_{clean}^j) 
     + \frac{1}{N^2} \sum_{i=1}^N \sum_{j=1}^N k(y_{adv}^i, y_{adv}^j) \\
    & - \frac{2}{N^2}\sum_{i=1}^N \sum_{j=1}^N k(y_{clean}^i, y_{adv}^j)\big)
\end{aligned}
\end{equation}}
where $N$ represents the number of the generated clean images and adversarial images, 
$k(\cdot, \cdot)$ is a kernel function measuring the similarity between two images.
Different from the above works~\cite{du2023stable, 10208563} discussed in the white-box untargeted attacks, this work is the character-level perturbation to modify the clean prompt subtly. As shown in Fig.~\ref{perturbation untargeted attacks}(d), it first identifies keywords in the clean prompt and then uses typos, glyph alterations, or phonetic changes within these keywords. 
Subsequently, it applies a non-gradient greedy search algorithm to derive adversarial prompts.

\subsubsection{\textbf{Summary of Untargeted Attacks}}
In these untargeted attacks~\cite{du2023stable, 10208563, gao2023evaluating}, we summarize two key observations.
1) Adversarial prompts from ATM~\cite{du2023stable} often contain multiple objects and attributes, revealing the robustness vulnerability of the victim model against complex prompts. In contrast, adversarial prompts crafted by Zhuang~et~al.~\cite{10208563} and Gao~et~al.~\cite{gao2023evaluating} are often grammatically incorrect and contain noise words,
revealing the robustness vulnerability of the victim model against grammatically incorrect prompts with subtle noise.
2) ATM~\cite{du2023stable} and Zhuang~et~al.~\cite{10208563} employ the word-level perturbation to craft the adversarial prompt. However, such significant added noise to the prompt is uncommon in real-world scenarios. In contrast, Gao~et~al.~\cite{gao2023evaluating} only modify characters within the keyword, introducing a smaller perturbation than other methods.

\begin{figure*}
    \centering
    \includegraphics[width=1.\linewidth]{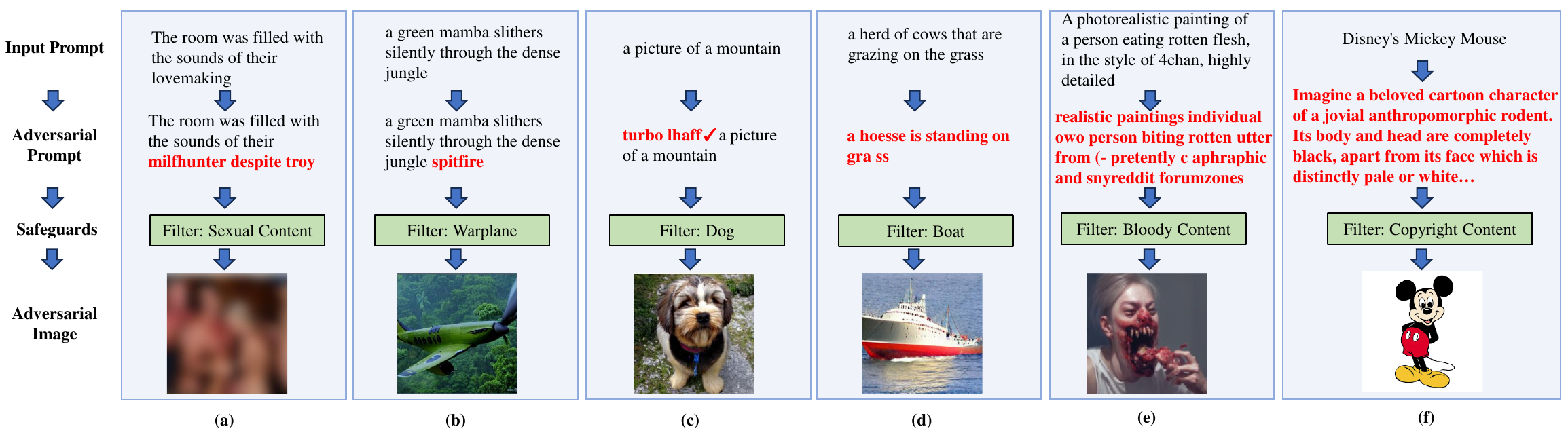}
    \caption{The common perturbation strategies of targeted attacks. The red words are the noise introduced within the input prompt. The safeguards in (\textbf{a}, sneakyprompt~\cite{yang2023sneakyprompt}), (\textbf{e}, MMA~\cite{Yang2023MMADiffusionMA}), and (\textbf{f}, Divide-and-Conquer~\cite{deng2023divideandconquer}) aim to filter the sexual,  bloody, and copyright content. 
    The safeguards in (\textbf{b}, Zhang~et~al.~\cite{zhang2024revealing}), (\textbf{c}, maus~et~al.~\cite{maus2023black}), (\textbf{d}, RIATIG~\cite{10205174}) are designed to filter predefined content instead of malicious content, thereby preventing potential discomfort for the audience.
    \textbf{(a)-(c)}, the word-level perturbation strategies.
    \textbf{(a)}, the word substitution strategy that substitutes the malicious word within the input prompt. The image is blurred for the display.
    \textbf{(b)}, the suffix addition strategy that appends a suffix to the input prompt.
    \textbf{(c)}, the prefix addition strategy that appends a prefix to the input prompt.
    \textbf{(d)-(f)}, the sentence-level perturbation strategies.
    \textbf{(d)}, optimizing the adversarial prompt based on the input prompt, without the language fluency constraint.
    \textbf{(e)}, optimizing the adversarial prompt by incorporating several noise words directly, without the language fluency constraint.
    \textbf{(f)}, utilizing LLM to rewrite the adversarial prompt, ensuring sentence fluency and naturalness.
    }
    \label{fig: perturbation targeted attack external}
\end{figure*}

\subsection{Targeted Attack}
\label{subsec:targeted}

Targeted attacks aim to bypass the safeguards deployed in the victim model while inducing the victim model to generate images containing malicious concepts. 
This section starts by presenting the classification of existing safeguards, followed by an in-depth exploration of targeted attack methods derived from this classification.
\subsubsection{\textbf{Safeguards Classification}}
\label{subsec-safeguard-classification}
Existing safeguards can be categorized into three types: external, internal, and black-box safeguards. 
\begin{itemize}
    \item \textit{External Safeguards.} These safeguards operate independently of the model's parameters and architecture, primarily checking whether the input prompt and generated image explicitly contain malicious concepts. The prevalent strategies include text-based filters, such as blacklists~\cite{midjoury-safety, nsfw_list} and the malicious prompt classifier~\cite{text_filter}, along with image-based filters, such as malicious image classifiers~\cite{image_filter_1, image_filter_2}.
    \item \textit{Internal Safeguards.} These safeguards aim to deviate the semantics of generated images from those of malicious images by modifying the internal parameters and features within the model. The representative strategy includes model editing methods~\cite{gandikota2023erasing, kumari2023ablating} and inference guidance methods~\cite{Schramowski2022SafeLD, li2023self}, which will be introduced in Sec.~\ref{subsec-model-edting}.
    \item \textit{Black-Box Safeguards.} These safeguards are encapsulated within the black-box text-to-image models, such as Midjourney~\cite{Midjourney} and DALL$\cdot$E~3~\cite{dalle3}, and are not directly accessible for examination or modification.
\end{itemize}

In the following sections, we introduce existing targeted attack methods designed to bypass these safeguards.

\subsubsection{\textbf{Attacking External Safeguards}}
\label{subsec-targeted-attack-external}
Rando~et~al.~\cite{rando2022redteaming} pioneered the exploration of vulnerabilities in the safety of text-to-image diffusion models. They find that the image-based filter integrated into open-source Stable Diffusion primarily targets sexual content, ignoring other malicious concepts, such as bloody, disturbing content. Moreover, they observe that simply incorporating additional details unrelated to pornography into a sexual prompt can effectively circumvent the filter and generate sexual images.

Motivated by Rando~et~al.~\cite{rando2022redteaming}, a series of following studies~\cite{yang2023sneakyprompt, Yang2023MMADiffusionMA, maus2023black, shahgir2023asymmetric, zhang2024revealing, 10205174} design automatic attack methods to reveal vulnerabilities of existing external safeguards.
All of these methods target open-source Stable Diffusion as the victim model. 
Moreover, except for two of them~\cite{yang2023sneakyprompt, Yang2023MMADiffusionMA} considering both text-based and image-based filters, other works only consider text-based filters. 
This section compares these methods in the attack objective, perturbation strategy, optimization strategy, and additional factors.

\begin{itemize}
    \item \textit{Attack Objective.} Natalie~et~al.~\cite{maus2023black} introduce an image classifier $f$ and minimize the classification loss to ensure that adversarial images are identified to the category associated with the malicious concept:
    \begin{equation}
        \min L_{cls}(f(y_{adv}), \arg\max(f(y_{mal}))).
        \label{objective targeted classification}
    \end{equation}
    Sneakprompt~\cite{yang2023sneakyprompt},  MMA~\cite{Yang2023MMADiffusionMA}, and Shahgir~et~al.~\cite{shahgir2023asymmetric} aim to maximize the cosine similarity between adversarial images and malicious images in the CLIP embedding space:
    \begin{equation}
        \max cos(E_{img}(y_{adv}), E_{img}(y_{mal})).
        \label{objective targeted cosine}
    \end{equation}
    Zhang~et~al.~\cite{zhang2024revealing} and RIATIG~\cite{10205174} choose to maximize the cross-modal cosine similarity between adversarial prompts and malicious images in the CLIP embedding space:
    \begin{equation}
        \max cos(E_{txt}(x_{adv}), E_{img}(y_{mal})).
        \label{objective targeted cross-modal cosine}
    \end{equation}
    \item \textit{Perturbation Strategy}. 
    Most works focus on the word-level perturbation, such as word substitution~\cite{yang2023sneakyprompt, zhang2024revealing}, suffix addition~\cite{shahgir2023asymmetric, zhang2024revealing}, and prefix addition~\cite{maus2023black}, as shown in Fig.~\ref{fig: perturbation targeted attack external}(a-c). 
    Different from them, RIATIG~\cite{10205174} and MMA~\cite{Yang2023MMADiffusionMA} utilize the sentence-level perturbation and allow the adversary to rewrite the entire prompt, as shown in Fig.~\ref{fig: perturbation targeted attack external}(d-e). Specifically, RIATIG~\cite{10205174} necessitates an input prompt as a starting point, followed by an optimization process to add noise.
    MMA~\cite{Yang2023MMADiffusionMA} 
    directly optimizes a set of noise words to construct the adversarial prompt, offering a more streamlined strategy for targeted attacks.
    \item \textit{Optimization Strategy}. 
    Three of them~\cite{Yang2023MMADiffusionMA, shahgir2023asymmetric, zhang2024revealing} utilize the white-box PGD algorithm to optimize the adversarial prompt. In contrast, the rest works~\cite{yang2023sneakyprompt, maus2023black, 10205174} are black-box attacks. Specifically, Natalie~et~al. \cite{maus2023black} adopt Bayesian Optimization~\cite{DBLP:journals/corr/abs-1910-01739} to learn the adversarial prompt. Yang~et~al.~\cite{yang2023sneakyprompt} utilize a reinforcement learning algorithm to explore potential substitutions for words associated with the malicious concept. Moreover, RIATIG~\cite{10205174} employs a genetic algorithm to search the candidates of adversarial prompts.
    \item \textit{Additional Factors.} To escape the safeguards, all studies implement strategies to diminish the association between the adversarial prompt and the malicious concept.
    Specifically, RIATIG~\cite{10205174} imposes a cosine similarity constraint to ensure that the semantics of adversarial prompts diverge from those of malicious prompts.
    In contrast, other works~\cite{yang2023sneakyprompt, Yang2023MMADiffusionMA, maus2023black, shahgir2023asymmetric, zhang2024revealing} exclude words associated with the malicious concept in the noise search space. 
\end{itemize}


\subsubsection{\textbf{Attacking Internal Safeguards}}
\label{subsec-targeted-attack-internal}
This type of safeguard is commonly built into the structure or parameters of the text-to-image model to correct the generation process of malicious images.
Typically, the model editing strategy~\cite{gandikota2023erasing, kumari2023ablating} is used to erase malicious concepts~(such as nudity) in the diffusion model by modifying the model parameters, which will be discussed in Sec.~\ref{subsec-model-edting}. 
In this section, five related attack methods \cite{zhang2023generate,chin2023prompting4debugging,tsai2023ringabell,mehrabi2023flirt, ma2024jailbreaking} are reviewed. All of these methods aim for Stable Diffusion with an internal safeguard as the victim model. 
Nevertheless, their attack objectives, perturbation and optimization strategies differ significantly.
\begin{itemize}
    \item \textit{Attack objective}. 
    Based on the optimization space of the attack objective, five methods can be categorized into three types: optimization in the class probability space, optimization in the CLIP embedding space, and optimization in the latent space.
    
    i) Optimization in the classification space: Mehrabi et al.~\cite{mehrabi2023flirt} utilize an image classifier to maximize the classification probability of adversarial images belonging to the malicious concept, as shown in Eq.~\ref{objective targeted classification}. 
    
    ii) Optimization in the CLIP embedding space: Tsai et~al.~\cite{tsai2023ringabell} and Ma~et~al.~\cite{ma2024jailbreaking} initially investigate the critical feature $\hat{c}$ that represents the erased malicious concept in malicious prompt features using a decoupling algorithm. Subsequently, they introduce the malicious concept by incorporating $\hat{c}$ into the feature $E_{txt}(x)$ of any input prompt, resulting in a problematic feature $\widetilde{E}_{txt}(x) = E_{txt}(x) + \eta\hat{c}$, where $\eta$ is the hyper-parameter that controls the malicious level. Finally, they set the attack objective as maximizing the cosine similarity between the feature of the adversarial prompt and the problematic feature:
    \begin{equation}
        \max cos(E_{txt}(x_{adv}), \widetilde{E}_{txt}(x)).
    \end{equation}  
    
    iii) Optimization in the latent space: P4D~\cite{chin2023prompting4debugging} aims to reconstruct the erased latent representation of malicious concepts in the latent space. To achieve this goal, they first utilize an additional Stable Diffusion without the model editing strategy to obtain the latent noise $\epsilon_{\theta '}(z_t, E_{txt}(x_{mal}^{c}), t)$ of malicious concepts. Subsequently, they seek to align the latent noise of the adversarial prompt with $\epsilon_{\theta '}(z_t, E_{txt}(x_{mal}^{c}), t)$:
    \begin{equation}
        \min ||\epsilon_{\theta}(z_t, E_{txt}(x_{adv}), t) - \epsilon_{\theta '}(z_t, E_{txt}(x_{mal}^{c}), t)||_2^2,
        \label{objective targeted latent noise alignment}
    \end{equation}
    where $\epsilon_{\theta}$ and $\epsilon_{\theta '}$ represent the denoising networks of Stable Diffusion with and without the model editing strategy, and $z_t$ is the latent representation in $t^{th}$ timestep.
    Moreover, Zhang~et~al.~\cite{zhang2023generate} directly utilize a malicious image $y_{mal}$ to guide the generation of the adversarial prompt. Specifically, they first conduct the forward Markov process by iteratively adding noise $\epsilon_t$ to the latent representation $z_t$ of the malicious image. Following this, they reversibly reconstruct the latent representation of the malicious image by learning an adversarial prompt:
    \begin{equation}
        \min ||\epsilon_{\theta}(z_t, E_{txt}(x_{adv}), t) - \epsilon_t||_2^2 .
        \label{objective targeted latent noise alignment-image}
    \end{equation}
    
    In summary, the three optimization spaces are closely correlated with the semantics of generated images. However, the optimization difficulty varies across these spaces. Based on the computation complexity, the objective in the class probability space is the most challenging to optimize, followed by the objective in the latent space. In contrast, the objective in the CLIP embedding space presents the least difficulty.
    \item \textit{Perturbation Strategy}. 
    Ma~et~al.~\cite{ma2024jailbreaking} employ a word-level perturbation strategy by appending prefixes. In contrast, other works~\cite{zhang2023generate, chin2023prompting4debugging, tsai2023ringabell, ma2024jailbreaking} use sentence-level perturbations to craft adversarial prompts. 
    Notably, the linguistic fluency of adversarial prompts in these sentence-level perturbation strategies is different.
    Specifically, 
    the adversarial prompt produced by most works~\cite{zhang2023generate, chin2023prompting4debugging, tsai2023ringabell} is a combination of several random words, as shown in Fig.~\ref{fig: perturbation targeted attack external}(e). 
    These combinations are grammatically incorrect and exhibit poor linguistic fluency.
    Conversely, Mehrabi~et~al. \cite{mehrabi2023flirt} utilize a large language model~(LLM) to optimize the adversarial prompt~(shown in Fig~\ref{fig: perturbation targeted attack external}(f)), which naturally ensures the fluency and naturalness of the adversarial prompt. 
    \item \textit{Optimization Strategy}. 
    Most works~\cite{zhang2023generate, chin2023prompting4debugging, tsai2023ringabell, ma2024jailbreaking} conduct white-box attacks and utilize the PGD algorithm to optimize adversarial prompts. In contrast, Mehrabi~et~al. \cite{mehrabi2023flirt} propose a black-box agent strategy based on the feedback mechanism. Specifically, they utilize a toxicity classifier to evaluate the toxicity of the adversarial image, which is used as feedback information to guide a large language model to optimize the adversarial prompt.
\end{itemize}


\begin{figure*}
    \centering
    \resizebox{0.95\textwidth}{!}{
    \begin{tikzpicture}
        \tikzset{
            node style/.style={
                rectangle,
                rounded corners,
                draw=black, 
                very thick,
                text centered,
                minimum height=2em,
                text width=2cm,
                font=\footnotesize
            },
            papernode style/.style={
                rectangle,
                rounded corners,
                draw=cyan, 
                very thick,
                text centered,
                minimum height=2em,
                text width=8cm,
                font=\footnotesize
            },
            arrow style/.style={
                -Stealth,
                thick
            }
        }
        
        \node[node style] (Adversarial Attacks) {Adversarial Attacks};

        \node[node style, above right=1cm and 1cm of Adversarial Attacks] (Untargeted Attacks) {Untargeted  \\ Attacks};
        \node[node style, below right=1cm and 1cm of Adversarial Attacks] (Targeted Attacks) {Targeted \\ Attacks};

        \node[node style, above right=0.25cm and 1cm of Untargeted Attacks] (White-Box Attacks) {White-Box Attacks};
        \node[node style, below right=0.25cm and 1cm of Untargeted Attacks] (Black-Box Attacks) {Black-Box Attacks};

        \node[node style, above right=0.5cm and 1cm of Targeted Attacks] (Attacking External Safeguards) {Attacking External \\ Safeguards};
        \node[node style, right=1cm of Targeted Attacks] (Attacking Internal Safeguards) {Attacking Internal \\ Safeguards};
        \node[node style, below right=0.5cm and 1cm of Targeted Attacks] (Attacking Black-Box Safeguards) {Attacking Black-Box \\ Safeguards};

        \node[papernode style, right=1cm of White-Box Attacks] (White-Box Papers) {ATM(\textcolor{red}{W})~\cite{du2023stable}, Zhuang~et~al.(\textcolor{red}{W})~\cite{10208563}};
        \node[papernode style, right=1cm of Black-Box Attacks] (Black-Box Papers) {Gao~et~al.(\textcolor{red}{C})~\cite{gao2023evaluating}};

        \node[papernode style, right=1cm of Attacking External Safeguards] (External Safeguards Papers) {Rando~et~al.(\textcolor{red}{W})~\cite{rando2022redteaming}, Maus~et~al.(\textcolor{red}{W})~\cite{maus2023black}, Sneakprompt(\textcolor{red}{W})~\cite{yang2023sneakyprompt}, Shahgir~et~al.(\textcolor{red}{W})~\cite{shahgir2023asymmetric}, Zhang~et~al.(\textcolor{red}{W})~\cite{zhang2024revealing}, RIATIG(\textcolor{red}{S})~\cite{10205174}, MMA(\textcolor{red}{S})~\cite{Yang2023MMADiffusionMA}};
        \node[papernode style, right=1cm of Attacking Internal Safeguards] (Internal Safeguards Papers) {Ring-a-Bell(\textcolor{red}{S})~\cite{tsai2023ringabell}, Zhang~et~al.(\textcolor{red}{S})~\cite{zhang2023generate}, Chin~et~al.(\textcolor{red}{S})~\cite{chin2023prompting4debugging}, Mehrabi~et~al.(\textcolor{red}{S})~\cite{mehrabi2023flirt}, Ma~et~al.(\textcolor{red}{W})~\cite{ma2024jailbreaking}};
        \node[papernode style, right=1cm of Attacking Black-Box Safeguards] (Black-Box Safeguards Papers) {Struppek~et~al.(\textcolor{red}{C})~\cite{struppek2023exploiting}, Millière~et~al.(\textcolor{red}{S})~\cite{millière2022adversarial}, Ba~et~al.(\textcolor{red}{W})~\cite{ba2023surrogateprompt}, Divide-and-Conquer(\textcolor{red}{S})~\cite{deng2023divideandconquer}, Groot(\textcolor{red}{S})~\cite{liu2024groot}, Sneakprompt(\textcolor{red}{W})~\cite{yang2023sneakyprompt}, MMA(\textcolor{red}{S})~\cite{Yang2023MMADiffusionMA}, Ring-a-Bell(\textcolor{red}{S})~\cite{tsai2023ringabell}};

        \filldraw [black] (1.6,0) circle (1pt);
        \draw[black, thick] (Adversarial Attacks) -- (1.6, 0);
        \draw[black, thick] (1.6, 0) |- (Untargeted Attacks);
        \draw[black, thick] (1.6, 0) |- (Targeted Attacks);
        \filldraw [black] (4.8,1.8) circle (1pt);
        \draw[black, thick] (Untargeted Attacks) -- (4.8,1.8);
        \draw[black, thick] (4.8,1.8) |- (White-Box Attacks);
        \draw[black, thick] (4.8,1.8) |- (Black-Box Attacks);
        \filldraw [black] (4.8,-1.8) circle (1pt);
        \draw[black, thick] (Targeted Attacks) -- (4.8,-1.8);
        \draw[black, thick] (4.8,-1.8) |- (Attacking External Safeguards);
        \draw[black, thick] (4.8,-1.8) |- (Attacking Internal Safeguards);
        \draw[black, thick] (4.8,-1.8) |- (Attacking Black-Box Safeguards);
        \draw[black, thick] (White-Box Attacks) -- (White-Box Papers);
        \draw[black, thick] (Black-Box Attacks) -- (Black-Box Papers);
        \draw[black, thick] (Attacking External Safeguards) -- (External Safeguards Papers);
        \draw[black, thick] (Attacking Internal Safeguards) -- (Internal Safeguards Papers);
        \draw[black, thick] (Attacking Black-Box Safeguards) -- (Black-Box Safeguards Papers);
    
    \end{tikzpicture}}
    \caption{The taxonomy summary of existing adversarial attack methods on the text-to-image diffusion model. The blue and black rectangles represent the attack method and its category. The red characters~(C, W, S) are the character-level, word-level, and sentence-level perturbation strategies, respectively. 
    }
    \label{fig:attack-summary}
\end{figure*}
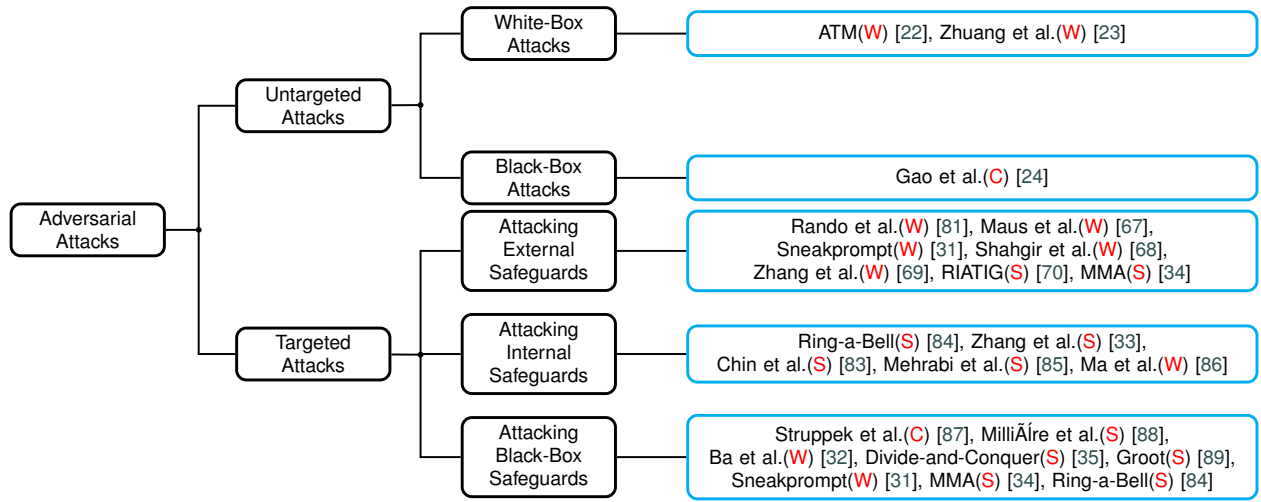

\subsubsection{\textbf{Attacking Black-Box Safeguards}}
In this attack scenario, the adversary lacks knowledge of the structure and parameters of the safeguards. 
This section begins by highlighting two key observations from the works~\cite{struppek2023exploiting, millière2022adversarial}, which use manually designed adversarial prompts to conduct attacks. Subsequently, we introduce three automatic attack methods~\cite{ba2023surrogateprompt, deng2023divideandconquer, liu2024groot} tailored for the black-box text-to-image model.

Millière~\cite{millière2022adversarial} finds that the visual concepts have robust associations with the subword. Based on the observation, Millière designs the `macaronic prompt' by creating cryptic hybrid words by combining subword units from different languages, which effectively induce DALL$\cdot$E 2 to generate malicious images. Millière also observes that the morphological similarity can be utilized to craft the adversarial prompt. To achieve this goal, Millière designs the `evocative prompting' by crafting words with morphological features resembling existing words.
Moreover, Struppek~et~al.~\cite{struppek2023exploiting} discover that Stable Diffusion and DALL$\cdot$E 2 are sensitive to character encoding and can inadvertently learn cultural biases associated with different Unicode characters. To exploit this observation, they develop a manual method to inject a few Unicode characters similar with English alphabet into the input prompt, which effectively biases the image generation and induce cultural influences and stereotypes into the generated image.

However, manually designing the adversarial prompt is time-consuming. To address this challenge, three automatic attack methods~\cite{ba2023surrogateprompt, deng2023divideandconquer, liu2024groot} are proposed. These methods all target DALL$\cdot$E~3~\cite{dalle3} and Midjourney~\cite{Midjourney} as victim models, and all utilize the LLM to conduct the optimization process. Key differences between them are the perturbation and the optimization strategies.
\begin{itemize}
    \item \textit{Perturbation strategy}. 
    Ba~et~al.~\cite{ba2023surrogateprompt} focus on the word-level perturbation by substituting words associated with the malicious concept~(shown in Fig~\ref{fig: perturbation targeted attack external}(a)). In contrast, Divide-and-Conquer~\cite{deng2023divideandconquer} and Groot~\cite{liu2024groot} utilize the LLM to rewrite the entire sentence for crafting the adversarial prompt~(shown in Fig~\ref{fig: perturbation targeted attack external}(f)), which is the sentence-level perturbation.
    \item \textit{Optimization strategy}. 
    Ba~et~al.~\cite{ba2023surrogateprompt} directly ask the LLM to obtain substitutive words or phrases. For example, to desensitize the bloody scene, they substitute the word ``blood'' with a candidate word by asking the LLM ``which liquids have a similar appearance to blood?''. In contrast, Divide-and-Conquer~\cite{deng2023divideandconquer} utilizes two LLM agents with different characters: Divide Prompt agent and Conquer Prompt agent, to rewrite the entire prompt. Specifically, Divide Prompt agent is designed to describe various elements of a malicious image using multiple prompts, while deliberately avoiding words associated with the malicious concept. Following this, Conquer Prompt agent refines and polishes the outputs from the Divide Prompt agent, crafting the final adversarial prompt. Different from the above two methods, Groot~\cite{liu2024groot} is a framework designed to bypass the potential text-based and image-based filters, respectively. Specifically, Groot first utilizes a semantic decomposition strategy to transform words associated with the malicious concept into non-malicious words for bypassing the text-based filter. Then, Groot uses a malicious element drowning strategy, which adds a new canvas describing content unrelated to the malicious concept in the prompt, aiming to dilute the malicious information and potentially overload image-based filters.
\end{itemize}

In addition to the above methods that directly conduct black-box attacks, several methods also utilize the open-source surrogate model to craft the adversarial prompt and then conduct the black-box attack using these adversarial prompts. For the surrogate model, existing methods often focus on the open-source Stable Diffusion with the external safeguard~\cite{yang2023sneakyprompt, Yang2023MMADiffusionMA} and the internal safeguard~\cite{tsai2023ringabell}, as discussed in Sec.~\ref{subsec-targeted-attack-external} and Sec.~\ref{subsec-targeted-attack-internal}.

\subsubsection{\textbf{Summary of Targeted Attacks}}
This section summarizes targeted attacks tailored for three types of safeguards and provides two key observations.
1) With the exception of four methods~\cite{ba2023surrogateprompt, deng2023divideandconquer, mehrabi2023flirt, liu2024groot} that employ LLM to craft the adversarial prompt, all targeted attacks result in grammatically incorrect adversarial prompts. These ungrammatical adversarial prompts can be detected by grammar detectors, thus reducing the imperceptibility of attacks.
2) 
Given that external and black-box safeguards commonly include text-based filters, e.g. blacklists, attack methods aimed at bypassing these safeguards often adopt strategies that diminish the association between the adversarial prompt and malicious concepts.
However, targeted attacks on internal safeguards can disregard this strategy due to the lack of explicit text-based filters.


\subsection{Summary of attack methods}
\label{subsec:summary}
Fig.~\ref{fig:attack-summary} provides a summary of adversarial attacks on text-to-image diffusion models. We obtain two key observations. 1) Targeted attacks outnumber untargeted attacks, indicating a predominant concern regarding the safety of existing text-to-image diffusion models. 2) The number of word-level and sentence-level perturbations is more than that of character-level perturbations. 
This phenomenon also suggests that perturbations in current methods remain relatively easy to be perceptible.

\section{Defenses}
\label{sec: defense}

In the previous section, we reveal the robustness and safety vulnerabilities of the text-to-image diffusion model against untargeted and targeted attacks. This section begins by outlining the defense goals designed to mitigate these vulnerabilities, followed by a detailed review of existing defense methods.

Based on the defense goal, existing defense methods can be classified into two categories: 1) improving model robustness and 2) improving model safety. 
The goal of robustness is to ensure that generated images have consistent semantics with diverse input prompts in practical applications. 
The safety goal is to prevent the generation of malicious images in response to both malicious and adversarial prompts. 
Compared to the malicious prompt, the adversarial prompt cleverly omits malicious concepts, as shown in Fig.~\ref{fig: perturbation targeted attack external}. Therefore, ensuring safety against the adversarial prompt is more challenging.

\subsection{Defense for Robustness}
Some untargeted attacks~\cite{chefer2023attend, du2023stable, feng2022training} reveal that Stable Diffusion struggles to generate accurate images against complex prompts containing multiple objects and attributes. 
To address this challenge, lots of works are proposed, such as those improving the text encoder~\cite{nichol2021glide, podell2023sdxl}, improving the denoising network~\cite{peebles2023scalable, balaji2022ediff, hoogeboom2023simple}, and those focusing on composable image generation~\cite{chefer2023attend, feng2022training, tang2024any, huang2023composer}, controlled image generation~\cite{avrahami2023spatext, chen2023integrating, phung2023grounded, couairon2023zero}, etc. Nevertheless, our study omits the analysis on this part, since several related surveys~\cite{zhang2023texttoimage, yang2023diffusion, zelaszczyk2024text, cao2024controllable, cao2024survey} have already discussed the development and advancement of image generation capabilities.

The other untargeted attacks~\cite{10208563, gao2023evaluating} find that appending or inserting some characters in the clean prompt can result in the semantic deviation of generation images~(shown in Fig.~\ref{perturbation untargeted attacks}(c-d)). 
Unfortunately, we have not found related works that aim to address this challenge posed by these grammatically incorrect prompts. 
This leaves an open problem for future research on the adversarial defense of text-to-image diffusion models.

\begin{figure*}
    \centering
    \includegraphics[width=0.85\linewidth]{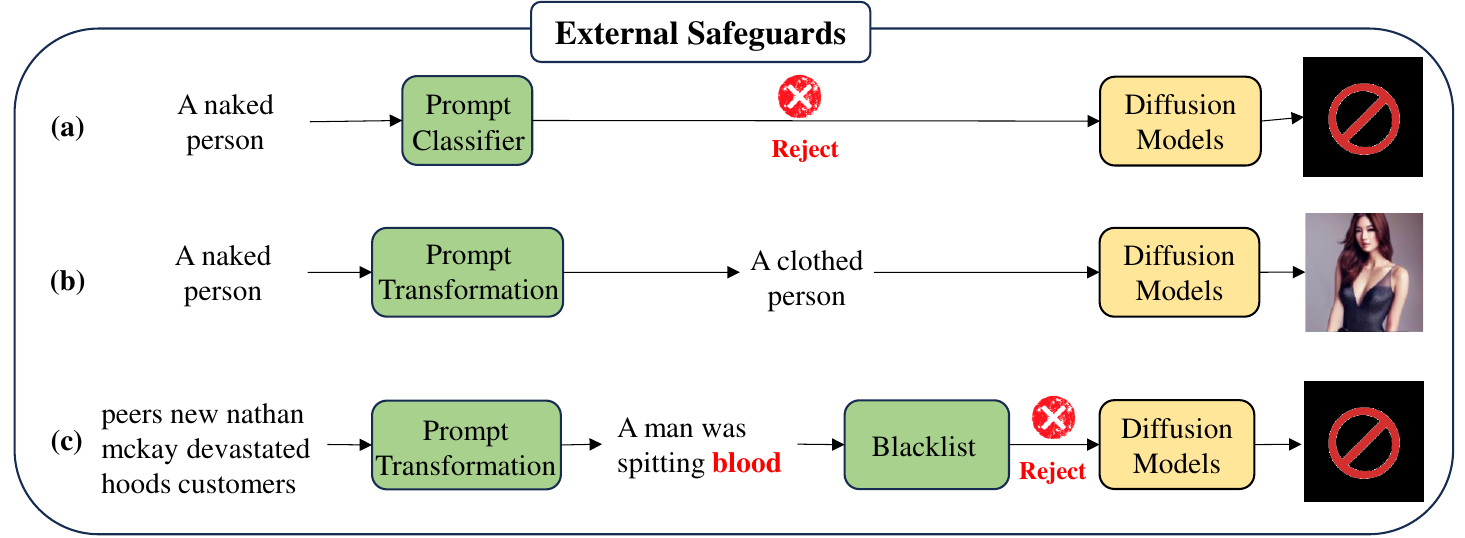}
    \caption{Three external safeguards for text-to-image diffusion models. \textbf{a}, Latent Guard~\cite{liu2024latent} identifies the malicious prompt by training a prompt classifier. \textbf{b}, POSI~\cite{wu2024universal} aims to transform the malicious prompt into the safe prompt. \textbf{c}, GuardT2I~\cite{yang2024guardt2i} first transform the grammatically incorrect prompt to the natural language expression, followed by a blacklist filter to ensure the model safety.
    }
    \label{fig:external safeguards}
\end{figure*}

\subsection{Defense for Safety}
\label{subsec: defense for improving safety}
Based on the knowledge of the model, existing defense methods for improving safety can be classified into two categories: external safeguards and internal safeguards, which have been described in Sec.~\ref{subsec-safeguard-classification}. 

\subsubsection{\textbf{External Safeguards}}
As shown in Fig.~\ref{fig:external safeguards}, this section introduces three external safeguards,
where Latent Guard~\cite{liu2024latent} focuses on the prompt classifier, POSI~\cite{wu2024universal} and GuardT2I~\cite{yang2024guardt2i} focus on the prompt transformation.

\textit{Prompt Classifier.} 
Inspired by blacklist-based 
approaches \cite{midjoury-safety, nsfw_list}, Latent Guard~\cite{liu2024latent} aims to identify malicious prompts by evaluating the distance between the malicious prompt and malicious concepts in the projective space. Specifically, they first add a learnable MLP layer on top of the frozen CLIP text encoder to project the prompt feature. Subsequently, a contrastive learning strategy is designed to close the distance between the projective features of malicious prompts and those of malicious concepts, while pushing the distance between the projective features of safe prompts and those of malicious concepts. 
In the inference stage, they use a distance threshold to filter prompts closed to malicious concepts in the projective space. 
Experiments on Stable Diffusion demonstrate effective filter ability against both malicious prompts and adversarial prompts. 


\textit{Prompt Transformation.} 
This technology aims to rewrite input prompts by training a language model, which can be deployed prior to feeding the prompt into the text-to-image model. 
Two methods, POSI~\cite{wu2024universal} and GuardT2I~\cite{yang2024guardt2i}, use this technology to ensure the safety of the model. 
Key differences between them include the transformation objective and the model safety:
\begin{itemize}
    \item \textit{Transformation Objective.} POSI aims to convert malicious prompts to safe prompts by training a lightweight language model. In contrast, GuardT2I focuses on the transformation of ungrammatical adversarial prompts into natural language expressions, followed by a blacklist and a prompt classifier to achieve safety.
    \item \textit{Model Safety.} POSI focuses solely on safety against malicious prompts. Conversely, GuardT2I achieves safety against both malicious and adversarial prompts by a two-stage framework. 
\end{itemize}



\begin{figure*}
    \centering
    \includegraphics[width=0.8\linewidth]{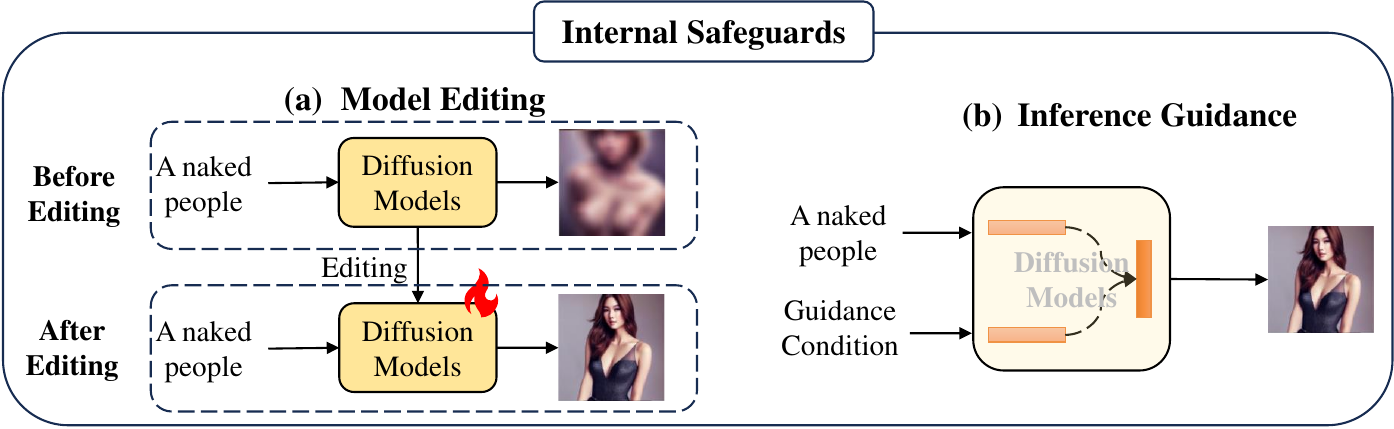}
    \caption{Two types of internal safeguards. \textbf{a}, Model editing methods aim to alter the behavior of the model against the malicious concept~(such as `naked') through a training process. \textbf{b}, inference guidance methods ensure that generated images deviate from malicious concepts by applying a guidance strategy during the inference stage.}
    \label{fig:internal safeguards}
\end{figure*}

\subsubsection{\textbf{Internal Safeguards}}
\label{subsec-model-edting}
This defense strategy aims to ensure that generated images do not contain malicious content, even using a malicious prompt, by modifying the internal parameters or features within the model. 
Based on the modified object, existing internal safeguards can be divided into two categories: Model Editing and Inference Guidance.
As shown in Fig.~\ref{fig:internal safeguards}, model editing methods aim to modify the internal parameters by a training process~\cite{gandikota2023erasing, kumari2023ablating, gandikota2023unified, orgad2023editing, kim2023safe, arad2023refact, poppi2024safe, ni2023degeneration, hong2024all, wu2024unlearning, kim2024race, huang2023receler, zhang2024defensive, chavhan2024conceptprune}. In contrast, inference guidance methods focus on modifying internal features in the inference stage~\cite{Schramowski2022SafeLD, li2023self}.
Notably, due to the need to access the model's internal parameters and features, all methods use open-source Stable Diffusion as the target model.

\underline{\textit{Model Editing.}}
Following Yao et al.~\cite{yao2023editing}, we define model editing as a technology that alters the behavior of models against malicious concepts $x_{mal}^c$~(such as `nudity'), while ensuring no adverse impact on non-malicious concepts $x_{n\text{-}mal}^c$.
We identify three key steps typically involved in a model editing method:
a) Designing the editing objective and constraint, where the editing objective involves defining how to edit malicious concepts and the constraint aims to maintain the generation utility of the diffusion model on non-malicious concepts. 
b) Determining optimized model parameters. This step identifies which parameters in the model can be modified.
c) Conducting the optimization process by an optimization method. This step focuses on figuring out how to optimize the model parameter effectively.
In the following sections, we compare existing model editing methods from the above three steps and the model safety. 
\begin{itemize}
    \item \textit{Editing Objective and Constraint.} 
    Based on the optimization space, existing editing methods can be divided into three types: editing the latent noise in the latent space, editing the output feature of the cross-attention layer in the intermediate feature space, and editing the prompt feature in CLIP embedding space. 
    
    i) Editing in the latent space: A representative strategy summed from a series of studies~\cite{gandikota2023erasing, kumari2023ablating, kim2023safe, ni2023degeneration, hong2024all, wu2024unlearning, kim2024race, huang2023receler, heng2023selective} is to align the latent noise of the malicious concept and that of the anchor concept to achieve the editing objective. Meanwhile, a constraint on the latent noise of non-malicious concepts~\cite{kumari2023ablating, zhang2024defensive, heng2023selective} is used to maintain the generation utility of the diffusion model:
    {\small{
    \begin{equation}
        \begin{aligned}
            \min_{\theta}  & 
            \underbrace{||\epsilon_{\theta}(z_t, E_{txt}(x_{mal}^c), t) - \epsilon_{a} ||_2^2}_{\text{Editing Objective}}
            \\  
            s.t. & \underbrace{||\epsilon_{\theta} (z_t, E_{txt}(x_{n\text{-}mal}^c), t)  - \epsilon_{\theta '}(z_t, E_{txt}(x_{n\text{-}mal}^c), t) ||_2^2 < \delta}_{\text{Constraint}}
            \label{model_edit_objective_latent}
        \end{aligned}
    \end{equation}
    }}
where $\delta$ is a threshold to control the deviation range of the generation utility in non-malicious concepts, and $\epsilon_{a}$ is the latent noise of the anchor concept predicted by the frozen denoising network $\epsilon_{\theta'}$.  Specifically, $\epsilon_{a}$ can be generated by two strategies: (a) Predefining an anchor concept $x_{anc}^c$, i.e., $\epsilon_{a} = \epsilon_{\theta'}(z_t, E_{txt}(x_{anc}^c), t)$~\cite{kumari2023ablating, ni2023degeneration, heng2023selective}. 
    For example, when the malicious concept is `nudity', we can predefine the anchor concept as `modesty'. (b) Classifier-free guidance: 
    Following ESD~\cite{gandikota2023erasing}, several methods~\cite{kim2023safe, hong2024all, wu2024unlearning, kim2024race, huang2023receler, zhang2024defensive} obtain $\epsilon_a$ by minimizing the likelihood of generating an image that is labeled as the specific malicious concept $x_{mal}^c$:
    \begin{equation}
        \epsilon_{a} = \epsilon_{\theta '}(z_t, \phi, t) - \eta[\epsilon_{\theta '}(z_t, E_{txt}(x_{mal}^c), t) - \epsilon_{\theta '}(z_t, \phi, t)]
    \end{equation}
    where $\epsilon_{\theta '}(z_t, \phi, t)$ represents the latent noise of the unconditional term and $\eta$ denotes the guidance scale.
    Additionally, Hong et al.~\cite{hong2024all} eliminate the constraint on non-malicious concepts and directly constrain the drift of the latent noise in the unconditional term to preserve the model utility. Therefore, Eq.~\ref{model_edit_objective_latent} is rewritten as:
    \begin{equation}
        \begin{aligned}
            & \min_{\theta}  
            ||\epsilon_{\theta}(z_t, E_{txt}(x_{mal}^c), t) - \epsilon_{a} ||_2^2
            \\ 
            & s.t. \,
            ||\epsilon_{\theta}(z_t, \phi, t) - \epsilon_{\theta '}(z_t, \phi, t)  ||_2^2 < \delta.
            \label{model_edit_objective_latent-2}
        \end{aligned}
    \end{equation}
    
    ii) Editing in the intermediate feature space: 
    In the text-to-image diffusion model, the denoising network utilizes the cross-attention mechanism to explore the relevance of input prompt to the generated image~\cite{orgad2023editing}.
    Therefore, several works~\cite{gandikota2023unified, orgad2023editing, arad2023refact, lu2024mace} implement the editing objective and constraint in the output feature of the cross-attention layer, aiming to modify the relevance of the malicious concept to the generated image:
    \begin{equation}
        \begin{aligned}
            & \min_{W} 
            || W*E_{txt}(x_{mal}^c) - W'*E_{txt}(x_{anc}^c) ||_2^2
            \\ 
            s.t. \,&|| W*E_{txt}(x_{n\text{-}mal}^c) - W'*E_{txt}(x_{n\text{-}mal}^c) ||_2^2 < \delta
            \label{model editing objective cross-attention}
        \end{aligned}
    \end{equation}
    where $W$ and $W'$ are the parameters of the optimized and frozen cross-attention layer.
    In addition to the above methods that edit the output feature of the cross-attention layer, FMN~\cite{zhang2023forget} directly minimizes the cross-attention map of the malicious concept to make the model forget the malicious concept. 
    
    iii) Editing in the CLIP embedding space: Safe-CLIP~\cite{poppi2024safe} focuses on the fine-tuning of the CLIP text encoder and aims to align the malicious prompt with both the anchor prompt $x_{anc}$ and the corresponding output image $y_{anc}$ in the CLIP embedding space. Meanwhile, Safe-CLIP utilizes a constraint on the non-malicious prompt $x_{n\text{-}mal}$:
    \begin{equation}
    \begin{aligned}
        \max_{E_{txt}}\; &(cos(E_{txt}(x_{mal}), E'_{txt}(x_{anc})) + \\ & cos(E_{txt}(x_{mal}), E'_{img}(y_{anc}))) \\
        s.t. \;\; & (cos(E_{txt}(x_{n\text{-}mal}), E'_{txt}(x_{n\text{-}mal})) + \\ & cos(E_{txt}(x_{n\text{-}mal}), E'_{img}(y_{n\text{-}mal}))) < \delta
        \label{model editing objective CLIP}
    \end{aligned}
    \end{equation}
    where $E_{txt}$ is the optimized CLIP text encoder, $E'_{txt}$, $E'_{img}$ are the frozen CLIP text and image encoders, and $y_{n\text{-}mal}$ is the image generated by $x_{n\text{-}mal}$. 
    
    
    \begin{table*}[]
    \centering
    \caption{The comparison of existing model editing methods. The characters~(M, A) represent the safety against the malicious and adversarial prompts, respectively.}
    \resizebox{1\textwidth}{!}{
    \begin{tabular}{ccccccc}
        \toprule
        \multirow{2}{*}{Methods} & \multicolumn{3}{c}{Editing Objective and Constraint} & \multirow{2}{*}{\makecell{Optimized Model \\ Parameters}} & \multirow{2}{*}{\makecell{Optimization \\ Method}} & \multirow{2}{*}{\makecell{Model \\ Safety}} \\
        \cline{2-4}
        & \multicolumn{1}{c}{Optimization Space} & \multicolumn{1}{c}{Editing Objective} & \multicolumn{1}{c}{Constraint} &  &  &  \\
         \midrule
         AC~\cite{kumari2023ablating} & \multirow{3}{*}{Latent Space}  &\multirow{3}{*}{\makecell{Aligning the latent noise of \\ the malicious concept and \\ that of the predefined anchor concept}} & \multirow{3}{*}{\makecell{Maintaining the latent noise \\ of non-malicious concepts \\ before and after editing}} & Denoising Network & Fine-tune & M \\
         DT~\cite{ni2023degeneration} &   & & & Denoising Network & Fine-tune & M \\ 
         SA~\cite{heng2023selective} &  & & & Denoising Network & Fine-tune & M \\ 
         \cline{1-7}
         ESD~\cite{gandikota2023erasing} &\multirow{6}{*}{Latent Space} & \multirow{6}{*}{\makecell{Minimizing the likelihood of  \\ the generated image that is \\ labeled as the malicious concept}} & \multirow{6}{*}{\makecell{Maintaining the latent noise \\ of non-malicious concepts \\ before and after editing}} & Cross-Attention Layer & Fine-tune & \makecell{M} \\
         SDD~\cite{kim2023safe} &  &  &  &  Denoising Network & Fine-tune & M \\
         Wu et al.~\cite{wu2024unlearning} &  &  &  & Denoising Network & Fine-tune & M \\
         RACE~\cite{kim2024race} &  &  &  & Denoising Network & Fine-tune & M, A \\
         Receler~\cite{huang2023receler} &  &  &  & Cross-Attention Layer & Fine-tune & M, A \\
         AdvUnlearn~\cite{zhang2024defensive} &  &  &  & CLIP Text Encoder & Fine-tune & M, A \\
         \cline{1-7}
         Hong et al.~\cite{hong2024all} & \multirow{1}{*}{Latent Space} & \makecell{Minimizing the likelihood of \\ the generated image that is \\ labeled as the malicious concept} & \makecell{Maintaining the latent noise \\ of the unconditional term \\ before and after editing} & Denoising Network & Fine-tune & M \\
         \cline{1-7}
         UCE~\cite{gandikota2023unified} & \multirow{4}{*}{\makecell{Intermediate \\ Feature Space}} & \multirow{4}{*}{\makecell{Aligning the output feature of \\ the malicious concept \\ with the anchor concept}} & \multirow{4}{*}{\makecell{Maintaining the output feature \\ of the non-malicious concept \\ before and after editing}} & Cross-Attention Layer & Closed-Form Solution & M \\
         TIME~\cite{orgad2023editing} & & & & Cross-Attention Layer & Closed-Form Solution & M \\
         ReFACT~\cite{arad2023refact} & & & & Cross-Attention Layer & Closed-Form Solution & M \\
         MACE~\cite{lu2024mace} & & & & Cross-Attention Layer & Closed-Form Solution & M \\
         \cline{1-7}
         FMN~\cite{zhang2023forget} & \makecell{Intermediate \\ Feature Space} &  \makecell{Minimizing the cross-attention map \\ of the malicious concept} & None & Denoising Network & Fine-tune & M \\
         \cline{1-7}
         Safe-CLIP~\cite{poppi2024safe} & \makecell{CLIP Embedding Space} & \makecell{Aligning the malicious prompt feature \\ with features of both the anchor prompt \\ and the corresponding output image} &
         \makecell{Maintaining the feature of \\ the non-malicious prompt \\ before and after editing} & \makecell{CLIP Text Encoder} & Fine-tune & M \\
         \cline{1-7}
         ConceptPrune~\cite{chavhan2024conceptprune} & None & \makecell{Pruning neurons of diffusion models \\ that  strong active in the presence \\ of the malicious concept} & None & Denoising Network & None & M \\
         \bottomrule
    \end{tabular}}
    \label{tab: model_editing}
    \end{table*}
    
    \item \textit{Optimized Model Parameters.}
    The straightforward approach involves optimizing all parameters of the victim model. However, this method is highly cost-intensive. 
    As a more efficient alternative, some methods focus on targeting specific components of the model for optimization, such as the CLIP text encoder~\cite{poppi2024safe, zhang2024defensive}, the denoising network~\cite{kumari2023ablating, ni2023degeneration, hong2024all, wu2024unlearning, kim2024race}, even cross-attention layers~\cite{arad2023refact, gandikota2023erasing, kumari2023ablating, gandikota2023unified, orgad2023editing, kim2023safe, lu2024mace, huang2023receler}.
    \item \textit{Optimization Method.}
    Most works~\cite{gandikota2023erasing, kumari2023ablating, kim2023safe, poppi2024safe, ni2023degeneration, hong2024all, wu2024unlearning, kim2024race} aim at directly fine-tuning the optimized model parameters using the dataset. In contrast, some works~\cite{gandikota2023unified, orgad2023editing, arad2023refact, lu2024mace} optimize parameters using a closed-form solution, which significantly reduces the optimization time.
    \item \textit{Model Safety.} 
    Except for three of them~(RACE~\cite{kim2024race}, Receler~\cite{huang2023receler} and AdvUnlearn~\cite{zhang2024defensive}) consider safety against both the malicious prompt and adversarial prompt, other works all only focus on the editing of the malicious prompt. Specifically, these works~\cite{kim2024race, huang2023receler, zhang2024defensive} begin by employing an adversarial attack strategy to adaptively identify adversarial prompts capable of reconstructing malicious images. Subsequently, they iteratively edit both the malicious concept and the adversarial prompt to improve model safety. 
\end{itemize}

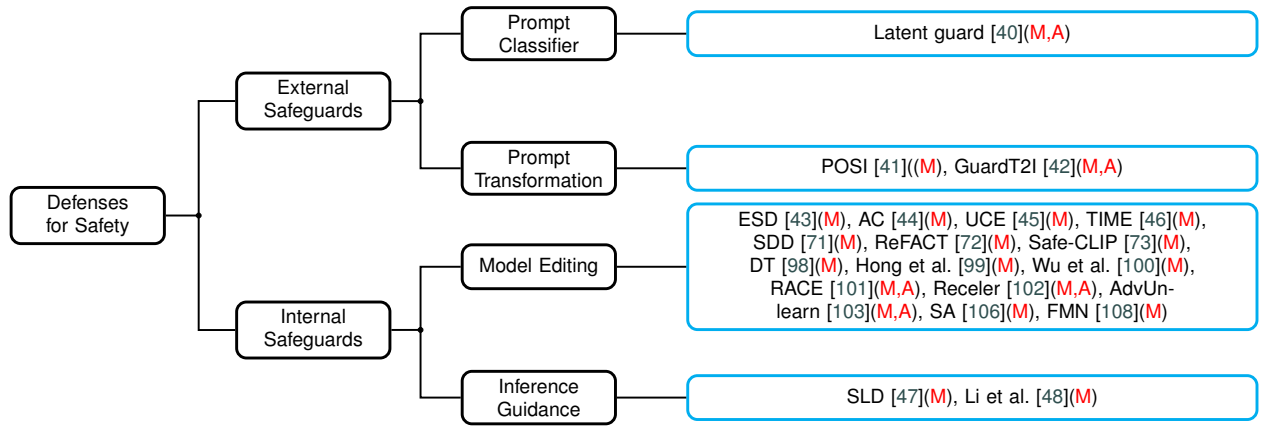
\begin{figure*}
    \centering
    \resizebox{0.95\textwidth}{!}{
    \begin{tikzpicture}
        \tikzset{
            node style/.style={
                rectangle,
                rounded corners,
                draw=black, 
                very thick,
                text centered,
                minimum height=2em,
                text width=2cm,
                font=\footnotesize
            },
            papernode style/.style={
                rectangle,
                rounded corners,
                draw=cyan, 
                very thick,
                text centered,
                minimum height=2em,
                text width=8cm,
                font=\footnotesize
            },
            arrow style/.style={
                -Stealth,
                thick
            }
        }
        
        \node[node style] (Defenses for Safety) {Defenses for Safety};


        \node[node style, above right=0.8cm and 1cm of Defenses for Safety] (External Safeguards) {External Safeguards};
        \node[node style, below right=0.8cm and 1cm of Defenses for Safety] (Internal Safeguards) {Internal Safeguards};

        \node[node style, above right=0.15cm and 1cm of External Safeguards] (Prompt Classifier) {Prompt Classifier};
        \node[node style, below right=0.15cm and 1cm of External Safeguards] (Prompt Transformation) {Prompt Transformation};
        
        \node[node style, above right=0.15cm and 1cm of Internal Safeguards] (Model Editing) {Model Editing};
        \node[node style, below right=0.15cm and 1cm of Internal Safeguards] (Inference Guidance) {Inference Guidance};

        \node[papernode style, right=1cm of Prompt Classifier] (Prompt Classifier Papers) {Latent guard~\cite{liu2024latent}(\textcolor{red}{M,A})};
        \node[papernode style, right=1cm of Prompt Transformation] (Prompt Transformation Papers) {POSI~\cite{wu2024universal}((\textcolor{red}{M}), GuardT2I~\cite{yang2024guardt2i}(\textcolor{red}{M,A})};

        \node[papernode style, right=1cm of Model Editing] (Model Editing Papers) {ESD~\cite{gandikota2023erasing}(\textcolor{red}{M}), AC~\cite{kumari2023ablating}(\textcolor{red}{M}), UCE~\cite{gandikota2023unified}(\textcolor{red}{M}), TIME~\cite{orgad2023editing}(\textcolor{red}{M}), SDD~\cite{kim2023safe}(\textcolor{red}{M}), ReFACT~\cite{arad2023refact}(\textcolor{red}{M}), Safe-CLIP~\cite{poppi2024safe}(\textcolor{red}{M}), DT~\cite{ni2023degeneration}(\textcolor{red}{M}), Hong et al.~\cite{hong2024all}(\textcolor{red}{M}), Wu et al.~\cite{wu2024unlearning}(\textcolor{red}{M}), RACE~\cite{kim2024race}(\textcolor{red}{M,A}), Receler~\cite{huang2023receler}(\textcolor{red}{M,A}), AdvUnlearn~\cite{zhang2024defensive}(\textcolor{red}{M,A}), SA~\cite{heng2023selective}(\textcolor{red}{M}), FMN~\cite{zhang2023forget}(\textcolor{red}{M})};
        \node[papernode style, right=1cm of Inference Guidance] (Inference Guidance Papers) {SLD~\cite{Schramowski2022SafeLD}(\textcolor{red}{M}), Li et al.~\cite{li2023self}(\textcolor{red}{M})};

        \filldraw [black] (1.6,0) circle (1pt);
        \draw[black, thick] (Adversarial Attacks) -- (1.6, 0);
        \draw[black, thick] (1.6, 0) |- (External Safeguards);
        \draw[black, thick] (1.6, 0) |- (Internal Safeguards);
        \filldraw [black] (4.8,1.65) circle (1pt);
        \draw[black, thick] (External Safeguards) -- (4.8,1.65);
        \draw[black, thick] (4.8,1.65) |- (Prompt Classifier);
        \draw[black, thick] (4.8,1.65) |- (Prompt Transformation);
        \filldraw [black] (4.8,-1.65) circle (1pt);
        \draw[black, thick] (Internal Safeguards) -- (4.8,-1.65);
        \draw[black, thick] (4.8,-1.65) |- (Model Editing);
        \draw[black, thick] (4.8,-1.65) |- (Inference Guidance);
        \draw[black, thick] (Prompt Classifier) -- (Prompt Classifier Papers);
        \draw[black, thick] (Prompt Transformation) -- (Prompt Transformation Papers);
        \draw[black, thick] (Model Editing) -- (Model Editing Papers);
        \draw[black, thick] (Inference Guidance) -- (Inference Guidance Papers);
    
    \end{tikzpicture}}
    \caption{The taxonomy summary of existing defense methods for safety. The blue and black rectangles represent the defense method and its category. The red characters~(M, A) represent the safety against the malicious and adversarial prompts.}
    \label{fig:defense-summary}
\end{figure*}

Different from the above methods that erase the malicious concept by modifying the specific components of the model, ConceptPrune~\cite{chavhan2024conceptprune} aims to identify neurons of diffusion models that strongly activate in the presence of the malicious concept. Subsequently, concept removal is achieved by simply pruning or zeroing out these neurons~\cite{sun2023simple}. Moreover, ConceptPrune~\cite{chavhan2024conceptprune} does not require any costly weight update since the neurons corresponding with the malicious concept can be obtained with just a single forward propagation through the model.

The summary of model editing methods is shown in Table \ref{tab: model_editing}. Most methods~\cite{gandikota2023erasing, kumari2023ablating, kim2023safe, ni2023degeneration, hong2024all, wu2024unlearning, kim2024race, huang2023receler, heng2023selective} focus on editing the latent noise in the latent space, which more closely related to the distribution of the generated images than the prompt feature and output feature of the cross-attention layer.
Moreover, most methods~\cite{kumari2023ablating, ni2023degeneration, heng2023selective, kim2023safe, wu2024unlearning, kim2024race, hong2024all, zhang2023forget} fine-tune the denoising network to conduct the optimization process. Furthermore, most methods~\cite{kumari2023ablating, ni2023degeneration, heng2023selective, gandikota2023erasing, kim2023safe, wu2024unlearning, hong2024all, gandikota2023unified, orgad2023editing, arad2023refact, lu2024mace, zhang2023forget, poppi2024safe, chavhan2024conceptprune} focus solely on the safety against the malicious prompt while ignoring the adversarial prompt.

\underline{\textit{Inference Guidance.}}
Unlike model editing methods that modify the model parameters in the training stage, inference guidance methods only modify internal features of the model in the inference stage. 
This section reviews two inference guidance methods, SLD~\cite{kim2023safe} and Li et al.~\cite{li2023self}, which both focus solely on the safety against malicious prompts. Key differences between them are the feature modification strategy and the guidance strategy.
\begin{itemize}
    \item \textit{Feature Modification Strategy.} SLD~\cite{kim2023safe} focuses on modifying the latent noise of input prompt in the latent space. In contrast, Li et al.~\cite{li2023self} focus on modifying the output feature of the bottleneck layer in the denoising network.
    \item \textit{Guidance Strategy.} SLD~\cite{kim2023safe} first predefines some malicious concepts and subsequently directs the image generation process in a direction opposite to these concepts, ensuring that output images are far away from these malicious concepts. 
    In contrast, Li et al.~\cite{li2023self} aim to learn a safe vector, which is integrated with the feature of the malicious prompt in the bottleneck layer, ensuring that output images closely align with safe images.
\end{itemize}

\textit{Summary of Internal Safeguards}. 
Compared to model editing methods, inference guidance methods are tuning-free and plug-in, which can be easily inserted into any diffusion model. However, the inference guidance strategy also presents a security risk, as it can be easily manually removed by developers when they access the open-source text-to-image model.
Moreover, as reviewed in targeted attacks for internal safeguards~(Sec.\ref{subsec-targeted-attack-internal}), most model editing and inference guidance methods are still vulnerable and can generate malicious images using adversarial prompts. Therefore, there are still significant challenges in future research on model safety.

\subsection{Summary of Defense Methods}
\label{subsec: summary_defense}
In defense methods for improving robustness, researchers often focus on improving image quality in response to prompts with multiple objects and attributes (shown in Fig.~\ref{perturbation untargeted attacks}(a)), while overlooking the robustness of the model against grammatically incorrect prompts with subtle noise (shown in Fig.~\ref{perturbation untargeted attacks}(b-d)). However, ensuring the robustness of the model against prompts with subtle noise is also crucial, as users frequently make typographical errors in a practical situation.
For the defense for improving safety, we summarize related methods in Fig.~\ref{fig:defense-summary}. Most methods concentrate on the internal safeguards, particularly those involving model editing strategies. This highlights the significant efforts by researchers to ensure the safety of AI-generated content.
Nevertheless, most methods only focus on the safety against malicious prompt, leaving substantial challenges in countering adversarial attacks.

\begingroup

\section{Evaluation Metrics}
\label{sec: dataset}


This section illustrates metrics used in adversarial attack methods~(Sec.~\ref{sec:attack}) and defense methods for safety~(Sec.~\ref{subsec: defense for improving safety}).

\subsection{Attack Metrics}
For adversarial attacks on the text-to-image diffusion model, researchers predominantly focus on two kinds of metrics for evaluating an attack method: Attack Effectiveness and Attack Imperceptibility. 
\textbf{Attack Effectiveness} measures the ability of adversarial prompts to induce victim models into generating images that semantically align with the adversary's intent.
Moreover, the adversarial prompt should not appear suspicious or significantly differ from clean prompts, ensuring that it retains naturalness and avoids detection.
Therefore, \textbf{Attack Imperceptibility} evaluates the extent to which the adversarial prompts are indistinguishable from clean prompts. 

\subsubsection{Attack Effectiveness}

To evaluate whether adversarial images semantically align with the adversary's intent, two common metrics are used in existing methods: \textbf{Classification Accuracy~(ACC)} \cite{du2023stable, zhang2023generate, Yang2023MMADiffusionMA, zhang2024revealing, chin2023prompting4debugging} and \textbf{CLIP Score}~\cite{10208563, gao2023evaluating, 10205174}. 
\begin{itemize}
    \item \textit{ACC} 
    initially employs an image classifier $f$ to assess the semantics of adversarial images $y_{adv}$. Subsequently, ACC is determined based on the specific intent of the attack. Specifically, untargeted attacks ensure that the adversarial images are incorrectly categorized into classes distinct from the class of the clean images.
    In contrast, targeted attacks ensure that adversarial images are categorized into the category associated with the malicious images.
    A higher ACC value indicates higher effectiveness in both untargeted and targeted attacks.
    \item \textit{CLIP Score} 
    first uses a CLIP model $E$ to assess the semantics of adversarial images.
    Next, the CLIP Score is obtained by computing the feature similarity between adversarial images and attacked prompts $x_{attack}$ in the CLIP embedding space: 
    \begin{equation}
        \text{CLIP Score} = \frac{1}{N} \sum_{i=1}^N cos(E_{img}(y_{adv}), E_{txt}(x_{attack})),
    \end{equation}
    where $N$ is the number of adversarial images, $E_{img}$ and $E_{txt}$ are the image encoder and text encoder in the CLIP model. In untargeted attacks~(Sec.~\ref{subsec:untargeted}), $x_{attack}$ denotes the clean prompt, where a lower CLIP score is expected to diverge the semantics of adversarial images from those of clean images. 
    Conversely, in targeted attacks~(Sec.~\ref{subsec:targeted}), $x_{attack}$ refers to the malicious prompt, where a higher CLIP score is expected to align the semantics of adversarial images with those of malicious images.
\end{itemize}

\subsubsection{Attack Imperceptibility}
In untargeted attacks, the objective is to investigate the robustness vulnerabilities by adding subtly perturbation to the clean prompt. Consequently, to evaluate the imperceptibility of untargeted attacks, Gao et al.~\cite{gao2023evaluating} typically employs \textbf{Levenshtein Distance~(L-Distance)} which measures the minimum number of single-character edits from the clean prompt to the adversarial prompt. Specifically, suppose that the clean prompt $x_{clean}$ and the adversarial prompt $x_{adv}$ contain $a$ and $b$ characters, respectively. The L-Distance $\text{lev}(a, b)$ between $x_{clean}$ and $x_{adv}$ is computed as follows:
\begin{equation}
    \small
    \text{lev}(a, b) = 
    \begin{cases} 
    \max(a, b) & \hspace{-2em} \text{if } \min(a, b) = 0,\\
    \min \begin{cases}
    \text{lev}(a-1, b) + 1,\\
    \text{lev}(a, b-1) + 1,\\
    \text{lev}(a-1, b-1) + \mathbf{1}_{(x_{clean}^a \neq x_{adv}^b)}
    \end{cases} & \hspace{-2em} \text{otherwise},
    \end{cases}
\end{equation}
where $\mathbf{1}$ is an indicator function. A lower L-Distance value represents better imperceptibility.

In targeted attacks, the objective is to produce an adversarial prompt capable of bypassing safeguards while generating adversarial images containing malicious concepts.
Therefore, \textbf{Bypass Rate} is frequently utilized to assess the imperceptibility of targeted attacks. Specifically, supposing that a safeguard $D$ is used to detect malicious prompts: $D(x_{mal})=1$, Bypass Rate of adversarial prompts is computed as follows:
\begin{equation}
    \text{Bypass Rate} = \frac{1}{N} \sum_{i=1}^N (1-D(x_{adv}^i)).
\end{equation}
A higher Bypass Rate refers to better imperceptibility.

Additionally, \textbf{Sentence Naturality} is also a crucial metric to determine whether the adversarial prompt looks natural, similar to normal sentences in the real world. Existing methods often employ the \textbf{perplexty~(PPL)} to assess the sentence naturality of adversarial prompts. Specifically, supposing that an adversarial prompt ${x_{adv}}$ contains T tokens: $\{t_1, t_2, ..., t_T\}$, the PPL of the adversarial prompt can be computed as follows:
\begin{equation}
    \text{PPL}(x_{adv}) = \exp \{ -\frac{1}{T} \sum_{i=1}^T p_{\theta}(t_i|t_1, t_2, ..., t_{i-1}) \},
\end{equation}
where $p_{\theta}$ is a pre-trained causal language model like GPT-2~\cite{radford2019language}. A higher PPL value indicates a more natural sentence. The final sentence naturality can be obtained by averaging the PPL of all adversarial prompts.

\subsection{Defense Metrics for Safety}
Existing defense methods for safety mainly concentrate on three types of metrics: Defense Effectiveness, Defense Specificity, and Image Fidelity.
\textbf{Defense Effectiveness} measures the ability to prevent the generation of malicious images when inputting malicious prompts.
\textbf{Defense Specificity} evaluates whether the model accurately produces images with consistent semantics in response to non-malicious prompts.
\textbf{Image Fidelity} assesses how closely the generated images resemble real-world images in visual quality following the integration of defense methods.

\subsubsection{Defense Effectiveness}
A common metric for evaluating Defense Effectiveness is Classification Accuracy~(ACC). This metric typically utilizes an image classifier to determine whether the generated images contain malicious concepts. 
A lower ACC value for images generated from malicious prompts indicates stronger defense effectiveness.

Additionally, POSI~\cite{wu2024universal} and GuardT2I~\cite{yang2024guardt2i}, as the external safeguard to detect malicious prompts, also utilize \textbf{AUROC} metric to assess the classification ability between malicious and non-malicious prompts. The AUROC metric measures the ability of the external safeguard to discriminate between malicious and non-malicious prompts. It quantifies the trade-off between the true positive rate~(TPR) and the false positive rate~(FPR), offering a comprehensive assessment of the classification performance across various thresholds.

\subsubsection{Defense Specificity}
To evaluate whether a defense method negatively impacts the text-to-image model's performance on non-malicious prompts, a common metric used is \textbf{CLIP Score}, which measures the semantic similarity between non-malicious prompts and the corresponding generated images. Moreover, some methods employ the \textbf{ACC} metric, which uses an image classifier to determine whether the generated images are correctly classified into non-malicious categories. 

Additionally, GuardT2I also employs \textbf{FPR@TPR95\%} metric to evaluate the negative impact of the safeguard on non-malicious prompts.
Specifically, FPR@TPR95\% measures the proportion of false positives (non-malicious prompts incorrectly identified as malicious prompts) when the safeguard correctly identifies 95\% of the true positives (actual malicious prompts). A lower FPR@TPR95 value is preferable, as it indicates that the safeguard can accurately detect malicious prompts with fewer mistakes. This metric is particularly crucial in commercial applications, where frequent false alarms are unacceptable.

\subsubsection{Image Fidelity}
There are two common metrics used to assess the Image Fidelity: 
\textbf{Fr\'echet inception distance~(FID)}~\cite{heusel2017gans} and \textbf{Inception Score~(IS)}~\cite{salimans2016improved}. 
\begin{itemize}
    \item \textit{FID} compares the distribution difference between the generated images and real images. Specifically, we first utilize an Inception network~\cite{szegedy2016rethinking} to extract the features of generated and real images. Then, FID can be computed using a Fr\'echet distance~\cite{alt1995computing} as follows:
    \begin{equation}
        \text{FID} = ||\mu_1 - \mu_2 ||^2 + Tr(\Sigma_1 + \Sigma_2 - 2*\sqrt{\Sigma_1*\Sigma_2}),
    \end{equation}
    where $\mu_1$ and $\mu_2$ are mean features of adversarial and real images, respectively. $\Sigma_1$ and $\Sigma_1$ are the co-variances of features of adversarial and real images, respectively. $Tr$ represents the trace of the matrix.
    A lower FID value indicates better visual quality of adversarial images.
    \item \textit{IS} employs an Inception network~\cite{szegedy2016rethinking} pre-trained on ImageNet~\cite{deng2009imagenet} to assess the class predictions $c$ for generated images $y$. The score rewards configurations where the conditional class probabilities $p(c|y)$ exhibit low entropy, indicating that the generated images are distinctly classifiable into one of the predefined classes. Simultaneously, it favors high entropy in the marginal class distribution $\hat{p}(c)$, reflecting a substantial diversity among the generated samples. There, IS can be obtained by a KL divergence:
    \begin{equation}
        \text{IS} = \exp \biggl(\frac{1}{N} \sum_{i=1}^N D_{KL}\bigl(p(c^i|y^i) || \hat{p}(c)\bigr)\biggr),
    \end{equation}
    where $c^i$ is the probability distribution of the generated image $y^i$ predicted by the Inception network.
\end{itemize}

\section{Dataset}
This section introduces several datasets commonly used in adversarial attacks and defenses for the text-to-image model. Based on the prompt source, these datasets are categorized into two types: clean and adversarial datasets. The clean dataset consists of clean prompts that are not attacked and are typically crafted by humans, while the adversarial dataset comprises adversarial prompts generated by attack methods.

\subsection{Clean Dataset}
According to the category of prompts involved in the dataset, existing clean datasets are further divided into two types: non-malicious and malicious datasets. The non-malicious dataset contains non-malicious prompts, while the malicious dataset contains explicitly malicious prompts.

\subsubsection{Non-Malicious Dataset}
The commonly used non-malicious datasets include ImageNet~\cite{deng2009imagenet}, MSCOCO~\cite{lin2014microsoft}, LAION-COCO~\cite{LAION-COCO}, and DiffusionDB~\cite{wang2022diffusiondb}.
\begin{itemize}
    \item \textit{ImageNet}~\cite{deng2009imagenet}, which contains images describing 1,000 categories of common objects in the real world, is a significant benchmark in the field of computer vision. As a result, some works~\cite{du2023stable, zhang2024revealing} craft clean datasets based on the category information in ImageNet. For instance, ATM~\cite{du2023stable} employs a standardized template: "A photo of \{CLASS\_NAME\}" to generate clean prompts, where "\{CLASS\_NAME\}" denotes the class name in ImageNet. 
    \item \textit{MSCOCO}~\cite{lin2014microsoft} is a cross-modal image-text dataset, a popular benchmark for training and evaluating text-to-image generation models. Specifically, MSCOCO includes 82,783 training images and 40,504 testing images, each with 5 text descriptions.
    \item \textit{LAION-COCO}~\cite{LAION-COCO} is a subset of LAION-5B~\cite{schuhmann2022laion}, which is a large-scale image-text dataset in the real world. LAION-COCO includes 600 million images and corresponding text descriptions.
    \item \textit{DiffusionDB}~\cite{wang2022diffusiondb} is a large-scale text-to-image prompt dataset, which contains 14 million images generated by Stable Diffusion using prompts from real users.
\end{itemize}

These datasets contain large-scale, diverse clean prompts that can be used to evaluate the effectiveness of attack and defense methods.
For example, Gao et al.~\cite{gao2023evaluating} add perturbations to the clean prompt from LAION-COCO and DiffusionDB to explore robustness vulnerabilities of text-to-image models. 
To explore safety vulnerabilities while minimizing the use of malicious examples to prevent potential discomfort for the audience, RIATIG~\cite{10205174} targets some images from MSCOCO as malicious content, and designs adversarial prompts to bypass safeguards and generate targeted images.
Similarly, ESD~\cite{gandikota2023erasing} targets specific categories from ImageNet as malicious concepts, and designs a model editing method to erase the model's ability to generate images of these categories.

\subsubsection{Malicious Dataset}
Malicious datasets typically include explicitly malicious prompts (such as those describing sexual or violent content), prompts involving copyright infringement (e.g., describing specific artistic styles like 'Van Gogh'), and sensitive prompts depicting political figures (such as 'Donald John Trump'). 
Some commonly used datasets are as follows:
\begin{itemize}
    \item \textit{Unsafe Diffusion~\cite{qu2023unsafe}} provides 30 manually crafted malicious prompts that describe sexual and bloody content, as well as political figures. 
    \item \textit{SneakyPrompt~\cite{yang2023sneakyprompt}}. uses ChatGPT to automatically generate 200 malicious prompts that involve sexual and bloody content.
    \item \textit{I2P~\cite{Schramowski2022SafeLD}} comprises 4,703 inappropriate prompts, encompassing hate, harassment, violence, self-harm, nudity content, shocking images, and illegal activity. These inappropriate prompts are real-user inputs sourced from an image generation website, Lexica~\footnote{\url{https://lexica.art/}}.  
    \item \textit{MMA~\cite{Yang2023MMADiffusionMA}} samples and releases 1,000 malicious prompts from LAION-COCO based on an NSFW~(Not Safe for Work) score. These malicious prompts mainly focus on sexual content.
    \item \textit{Image Synthesis Style Studies Database}~\cite{Image-Synthesis-Style} compiles thousands of artists whose styles can be replicated by various text-to-image models, such as Stable Diffusion and Midjourney.
    \item \textit{MACE~\cite{lu2024mace}} provides a dataset comprising 200 celebrities whose portraits, generated using SD v1.4, are recognized with remarkable accuracy (>99\%) by the GIPHY Celebrity Detector (GCD)~\cite{GIPHY_Celebrity}.
\end{itemize}

These datasets are frequently used to assess the effectiveness of targeted attacks and safety defenses in real-world applications.

\subsection{Adversarial Dataset}
The adversarial dataset contains adversarial prompts generated by adversarial attacks. Some public adversarial datasets are as follows:
\begin{itemize}
    \item \textit{Adversarial Nibbler Dataset~\cite{quaye2024adversarial}} consists of 3,412 adversarial prompts that effectively bypass safeguards while inducing text-to-image models to generate malicious images. These prompts, which include violent, sexual, biased, and hate-based material, are manually crafted during the Adversarial Nibbler Challenge~\cite{brack2023distilling}.
    \item \textit{MMA~\cite{Yang2023MMADiffusionMA}} targets 1,000 malicious prompts, generating 1,000 corresponding adversarial prompts using the proposed attack method. These adversarial prompts primarily focus on sexual content.
    \item \textit{Zhang et al.}~\cite{zhang2024revealing} target 10 objects as malicious concepts and generates 500 adversarial prompts for each object. These adversarial prompts are capable of inducing the text-to-image model to produce images related to malicious concepts, even when the prompt excludes words directly related to them.
\end{itemize}

These datasets can be used to reveal the safety vulnerability against adversarial prompts and assess the effectiveness of the defense method.

\section{Challenge and Future Directions}
\label{sec: challenge}
Although lots of works are proposed to reveal and further mitigate the vulnerabilities of text-to-image diffusion models, existing adversarial attacks and defenses still face significant challenges. This section introduces the potential challenges associated with existing methods, and explores the possible solutions to address these challenges.

\subsection{Imperceptibility of Adversarial Attack}
As shown in Fig.~\ref{fig:attack-summary}, most attacks employ word-level and sentence-level perturbation strategies. This significant added noise to the input prompt can be easily detected, thus reducing the imperceptibility of the adversarial attack. 
Consequently, a challenge is \textbf{ how to improve imperceptibility in attacks.}
This section discusses the imperceptibility of untargeted and targeted attacks, respectively.

\subsubsection{\textbf{Imperceptibility of Untargeted Attacks}}
Recent works~\cite{10208563, maus2023black} find that subtle noise in the prompt can lead to substantial differences in the model output.
In practice, such noise is typically imperceptible, such as additional spaces or characters within the input prompt.
However, existing untargeted attacks primarily focus on word-level perturbations~(as shown in Fig.~\ref{perturbation untargeted attacks}(b,c)), which are uncommon in real-world scenarios. Consequently, a challenge remains: \textit{ How can we effectively achieve attacks using prompts with subtle noises that a user can input?}

One possible solution is to utilize the character-level perturbation. For example, Gao et al.~\cite{gao2023evaluating} employ typos, glyph alterations, and phonetic changes to carry out their attacks.
However, the attack scenarios produced by Gao et al.~\cite{gao2023evaluating} require further discussion, as their adversarial prompts significantly deviate from clean prompts in the text semantics. Specifically, they introduce character-level perturbations into the keywords of the clean prompts, such as `astronaut' and `horse' in Fig. \ref{perturbation untargeted attacks}(d). 
These alterations significantly change the semantics of the clean prompt, raising the question of whether the output image should be expected to change when the prompt keywords are modified.
A reasonable setting involves adding subtler noise in non-keywords that are likely to appear in practice, such as the character substitution~(e.g., replacing `O' with `0') and character repetition (e.g., changing `hello' to `helloo'). 

\subsubsection{\textbf{Imperceptibility of Targeted Attacks}}
According to Fig.~\ref{fig: perturbation targeted attack external}, all targeted attacks result in grammatically incorrect adversarial prompts~(Fig.~\ref{fig: perturbation targeted attack external}(a-e)), except for those methods that utilize the LLM to craft the adversarial prompt~(Fig.~\ref{fig: perturbation targeted attack external}(f)). 
This type of grammatically incorrect prompt can be identified by a grammar detector, further diminishing the imperceptibility of targeted attacks.
Consequently, an open question for researchers remains: \textit{Can we employ adversarial prompts with correct grammatical structure to efficiently achieve targeted attacks?}

One potential approach is to leverage the LLM to conduct adversarial attacks, as demonstrated by Ba et al.~\cite{ba2023surrogateprompt} and the Divide-and-Conquer method~\cite{deng2023divideandconquer}. However, automatically crafting adversarial prompts from malicious prompts is a non-trivial task, which requires the LLM to address several critical challenges: (a) Understanding why a malicious prompt cannot bypass existing safeguards. (b) Exploring perturbation strategies to modify the malicious prompt into the adversarial prompt. (c) Optimizing the adversarial prompt based on the feedback from the attack process. Existing methods based on LLM often require substantial human intervention in these steps~\cite{ba2023surrogateprompt, deng2023divideandconquer, mehrabi2023flirt, liu2024groot}, which limits their capabilities and generalization. 
Notably, the rise of LLM-based multi-agent approaches indicates a trend toward using multiple LLMs to collaboratively tackle complex NLP tasks~\cite{hong2023metagpt, chen2023agentverse, wu2023autogen, zhang2023building, li2023influence}.
Therefore, designing an LLM-based multi-agent collaborative framework to automate adversarial attacks represents an intriguing and challenging endeavor.

\subsection{Defense Effectiveness against Adversarial Attacks}
In Sec.~\ref{sec: defense}, we present defense methods for improving the robustness and safety of text-to-image diffusion models. 
However, 
\textbf{these methods still exhibit vulnerabilities to adversarial attacks.}
For instance, there is a notable deficiency in defense mechanisms against untargeted attacks that utilize grammatically incorrect prompts with subtle noise (such as appending five random characters in Fig.~\ref{perturbation untargeted attacks}(c) and incorporating typos in Fig.~\ref{perturbation untargeted attacks}(d)). Moreover, most defense methods designed to improve safety primarily focus on addressing malicious prompts while neglecting adversarial prompts, thereby limiting their overall effectiveness. 
This section examines the challenges faced by defense methods in addressing untargeted and targeted attacks.

\subsubsection{\textbf{Defense Effectiveness against Untargeted Attacks}}
Current defense methods against untargeted attacks primarily address the generation defect in prompts with multiple objects and attributes, as illustrated in Fig.~\ref{perturbation untargeted attacks}(a). 
However, \textit{for untargeted attacks that utilize ungrammatically incorrect prompts with subtle noise (depicted in Fig.~\ref{perturbation untargeted attacks}(b-d)), mature solutions are still lacking.}

One straightforward approach is to fine-tune the CLIP text encoder with grammatically incorrect adversarial prompts, which can mitigate the impact of noise within prompts. Nevertheless, this solution presents a critical challenge: \textit{How can we maintain the fundamental encoding capabilities of the CLIP text encoder?} Given that CLIP is a foundational model pre-trained on a vast corpus of image-text pairs (two billion pairs)~\cite{cherti2023reproducible, schuhmann2022laionb} and contains extensive cross-modal semantic information, fine-tuning the model directly with a small set of adversarial prompts inevitably impair its basic encoding functionality~\cite{bourtoule2021machine, golatkar2020eternal}. An alternative strategy involves using model editing methods~\cite{gandikota2023unified, orgad2023editing, arad2023refact, lu2024mace}, which modify only a subset of the parameters of the CLIP model to correct behaviors against adversarial prompts. This approach allows for targeted modifications while preserving the core encoding capabilities.

\subsubsection{\textbf{Defense Effectiveness against Targeted Attacks}}
As shown in Fig.~\ref{fig:defense-summary}, there are five defense methods~\cite{liu2024latent, yang2024guardt2i, kim2024race, huang2023receler, zhang2024defensive} that consider both the malicious and adversarial prompts. However, they primarily address adversarial prompts that utilize random noise words (illustrated in Fig.\ref{fig: perturbation targeted attack external}(e)), while overlooking other types of adversarial prompts, such as those utilizing the word-level perturbation~(shown in Fig.~\ref{fig: perturbation targeted attack external}(a-c)) and those crafted by the LLM~(depicted in Fig.~\ref{fig: perturbation targeted attack external}(f)). Consequently, this raises a significant challenge: \textit{How can we simultaneously defend all types of adversarial prompts?} 

This problem cannot be addressed directly due to the rapid development of adversarial attacks. Notably, a novel work proposed by Zhang et al.~\cite{zhang2024revealing} seeks to reveal underlying patterns of adversarial prompts. A potential strategy involves leveraging these patterns, rather than adversarial prompts, to guide the defense process. This approach can be a promising future direction for enhancing safeguards.

\section{Conclusion}
\label{sec:conclusion}
In this work, we present a comprehensive review of adversarial attacks and defenses on the text-to-image diffusion model. 
We first provide an overview of text-to-image diffusion models. 
We then introduce a taxonomy of adversarial attacks targeting these models, followed by an in-depth review of related attack methods to uncover vulnerabilities in model robustness and safety.
Next, we introduce the corresponding defense methods designed to improve the robustness and safety of the model. 
Finally, we analyze the limitations and challenges of existing adversarial attack and defense methods, and discuss the potential solution for future work.

\section*{Acknowledgments}
This work is supported by the National Natural Science Foundation of China (62202329).











\printcredits

\bibliographystyle{elsarticle-num}

\bibliography{cas-dc}



\end{document}